\documentclass[10pt,aps,prd,onecolumn,showpacs,amsmath,amssymb,nofootinbib,eqsecnum,preprintnumbers,superscriptaddress]{revtex4-2}

\usepackage{mathtools}
\usepackage{esvect}
\usepackage{appendix}
\usepackage{tikz}
\usepackage{changepage}
\usepackage{xcolor}
\usepackage{amsfonts}
\usepackage{longtable}
\usepackage{physics}
\usepackage{xspace}
\usepackage{mathtools}
\usepackage{amsmath}
\usepackage{amsthm}
\usepackage{tikz}
\usepackage{makecell}
\usepackage{comment}
\usepackage{multirow}
\usepackage{titlesec}
\usepackage[colorlinks=true,
            citecolor=red,
            linkcolor=blue,
            urlcolor=violet,
            filecolor=cyan,
            backref=false]{hyperref}
\allowdisplaybreaks
\DeclareSymbolFont{matha}{OML}{txmi}{m}{it}
\DeclareMathSymbol{\varv}{\mathord}{matha}{118}

\titleformat{\paragraph}[block]{\filcenter}{}{0pt}{}

\DeclareMathSymbol{*}{\mathbin}{symbols}{"03} 
\DeclareMathSymbol{\ast}{\mathbin}{symbols}{"03}

\newcommand{\lie}{\pounds}

\renewcommand{\cosh}{\operatorname{ch}}
\renewcommand{\sinh}{\operatorname{sh}}
\renewcommand{\tanh}{\operatorname{th}}
\renewcommand{\coth}{\operatorname{cth}}
\renewcommand{\csch}{\operatorname{csch}}
\renewcommand{\sech}{\operatorname{sech}}

\begin{document}


\title{Canonical quantization of all minisuperspaces with consistent symmetry reductions}

\begin{abstract}
    We present the quantization of all symmetry reductions of the Einstein--Hilbert Lagrangian that correctly reproduce the reduced Einstein's field equations --- i.e., characterized by the infinitesimal group actions obeying the principle of symmetric criticality. These correspond to the spacetime symmetries of spherical/hyperbolic/planar Schwarzschild/Taub--NUT, BI/BII/BIII-metrics, near-horizon extreme Kerr geometry, swirling universe, closed/open/flat FLRW cosmologies, other FLRW-type metrics, and Bianchi type I, II, VIII, and IX spacetimes. We derive the Hamiltonian and the conformal symmetries of the superspace metrics (the conditional symmetries), promote them to operators, and solve the Wheeler--DeWitt equation with and without imposing these symmetries.
\end{abstract}

\author{Poula Tadros}

\email{poula.tadros@matfyz.cuni.cz}

\affiliation{Institute of Theoretical Physics, Faculty of Mathematics and Physics,
Charles University, Prague, V Hole{\v s}ovi{\v c}k{\' a}ch 2, 180 00 Prague 8, Czech Republic}

\author{Ivan Kol{\'a}{\v r}}

\email{ivan.kolar@matfyz.cuni.cz}

\affiliation{Institute of Theoretical Physics, Faculty of Mathematics and Physics,
Charles University, Prague, V Hole{\v s}ovi{\v c}k{\' a}ch 2, 180 00 Prague 8, Czech Republic}

\author{Otakar Sv\'itek}

\email{otakar.svitek@matfyz.cuni.cz}

\affiliation{Institute of Theoretical Physics, Faculty of Mathematics and Physics,
Charles University, Prague, V Hole{\v s}ovi{\v c}k{\' a}ch 2, 180 00 Prague 8, Czech Republic}

\maketitle
\section{Introduction}

Quantizing gravity is one of the central challenges in theoretical physics. The quest for quantizing gravity began in 1930s, where an attempt to derive a consistent quantum theory from linearized gravity was made (see \cite{Bronstein:2012zz} for an English translation of the original paper). However, the problem that halted this attempt was the realization that the minimal length built into quantum mechanics (Planck's length) is incompatible with dynamical spacetimes. Other attempts were made by treating gravity as a quantum field theory of spin $2$ particles (gravitons) on a fixed background \cite{Feynman:1963ax,PhysRev.135.B1049}; however, it was shown to be non-renormalizable in \cite{Goroff:1985sz}. An effective field theory (EFT) approach was proposed to address the non-renormalizability problem \cite{Weinberg:1980gg,Donoghue:1994dn,Burgess:2003jk}, but it does not provide an ultraviolet-complete theory, leaving the quantization incomplete (see \cite{Burgess:2020tbq} for a review). Other approaches include gravity on non-commutative manifolds \cite{Snyder:1946qz,CONNES199129,Chamseddine:1996zu} (see \cite{Aastrup:2012jj} for a review), causal sets \cite{1991regr.conf..150S,Bombelli:1987aa} (see \cite{Surya:2019ndm} for a review), loop quantization \cite{Rovelli:1997yv,Rovelli:2010bf,Ashtekar:2007px,Ashtekar:2007tv,Smolin:2004sx,Ashtekar:1986yd} (for reviews see \cite{Rosas-Rodriguez:2005esi,BarberoG:1995abd} and applications \cite{Tibrewala:2015eta,Sartini:2021ktb,Agullo:2016tjh,Zhu:2017jew,Corichi:2016xfy,Bojowald:2005epg,Bojowald:2004ax,Bojowald:2008ma,BarberoG:2010oga}), as well as more radical proposals such as string theory \cite{Veneziano:1968yb,Nambu:1986ze,Goto:1971ce,GREEN1984117}, each with its own limitations; see \cite{Demulder:2023bux} for a review.

For this paper, we focus on canonical quantization, introduced in \cite{PhysRev.160.1113}, where general relativity (GR) was quantized by promoting the induced metric and its conjugate momentum to operators and solving the Hamiltonian constraint, known as the \textit{Wheeler–DeWitt (WDW) equation}. However, the resulting functional equation is generally ill-defined, since it is formally equivalent to an infinite set of differential equations. This difficulty can be avoided in highly symmetric spacetimes by reducing the theory to a theory on \textit{minisuperspaces} or \textit{midisuperspaces}, i.e. the spaces of metrics sharing the same symmetries. This is achieved by inserting a suitable metric ansatz directly into the Lagrangian. If all functions in the metric ansatz depend on a single variable, the superspace is said to be truncated to a minisuperspace. By doing so, a system with an infinite number of degrees of freedom is reduced to a finite-dimensional one, thus resembling the quantization of a particle rather than a field theory. This type of quantization is the main focus of the present paper. A more field-theoretic quantization arises when the functions in the metric ansatz used to reduce the superspace depend on multiple variables. This leads to a system with infinitely many degrees of freedom, but still more manageable than the full superspace, and is referred to as a midisuperspace. Midisuperspace quantization has been used to canonically quantize static spherically symmetric spacetimes in \cite{Kuchar:1994zk,Berger:1972pg}\footnote{While static spherically symmetric spacetimes truncate to a minisuperspace, midisuperspace techniques are also applicable.} and spacetimes with cylindrical or toroidal symmetries in \cite{Torre:1998dy}.

For minisuperspaces, canonical quantization of static spherically symmetric metrics was performed in \cite{Christodoulakis:2012eg,Melas:2013hoa,Christodoulakis:2013xja,Melas:2015rwa}, with charged extensions in \cite{Christodoulakis:2013sya,Dimakis:2017qcf,Melas:2015gqa} and supersymmetric generalizations in \cite{Pollari:2025hui}. FLRW minisuperspaces coupled to scalar fields were introduced and quantized in \cite{Zampeli:2015hba,Geiller:2022baq,Zampeli:2018tye,Svitek:2016uda,Christodoulakis:2018swq,Robles-Perez:2018kcq,Paliathanasis:2015ffn}, and further extended to include gauge fields in \cite{Zhang:2015aaa,Vakili:2014sda,Paliathanasis:2015ffn}. Bianchi models were treated in \cite{Geiller:2022baq,Christodoulakis:2004nw}, including scalar-field couplings in \cite{Zampeli:2015ojr,Paliathanasis:2016rho}, the locally rotationally symmetric (LRS) Bianchi III case in \cite{Karagiorgos:2017nta}, and their semiclassical analysis in \cite{Zampeli:2015ojr}, with a review in \cite{Faraoni:2021opj}. FLRW cosmologies in scalar-tensor and Brans--Dicke theories were quantized in \cite{Borowiec:2021tyg,Paliathanasis:2019luv}. In $(2+1)$ dimensions, the quantization was carried out for the BTZ black hole \cite{Christodoulakis:2014wba} and homogeneous string cosmologies \cite{Naderi:2020wat}. Minisuperspace quantization has also been widely applied in modified gravity, including $f(T)$ \cite{Dimakis:2023oje}, $f(T,B)$ \cite{Paliathanasis:2021kuh}, $f(R)$, $f(T)$ and $f(\mathcal{G})$ \cite{Capozziello:2022vyd,Paliathanasis:2019ega}, $f(Q)$ \cite{Bajardi:2023vcc,Dimakis:2021gby,De:2025swf}, $f(R)$ with perfect fluids \cite{Dimakis:2013oza}, and Einstein–aether theory \cite{Roumeliotis:2018ook}, as well as bubbling $AdS_2 \times S^2$ geometries in supergravity \cite{Li:2016xni} and Bianchi I, III, and Kantowski--Sachs models in teleparallel and $f(T,B)$ gravity \cite{Paliathanasis:2022vux,Paliathanasis:2022mcc,Paliathanasis:2022mrp}, with further results in $f(R)$ Bianchi cosmologies \cite{Obaidullah:2022kad,Obaidullah:2022dgc} and additional cosmological applications reviewed in \cite{Ida:2013sra,Ashtekar:1993wb,Oriti:2024qav,Kung:1993xi,Gryb:2018whn,Manojlovic:1993tr,Kan:2021fmw,Rajeev:2021lqk,Rajeev:2021yyl,Mondal:2023cxx}.

The very first step in the quantization procedure is to derive the reduced Lagrangian by substituting the spacetime metric ansatz into the theory's Lagrangian and contracting with the appropriate $l$-chain (to reduce the volume element on the manifold to the volume element on the reduced space). For minisuperspaces and for the Einstein--Hilbert Lagrangian, the reduced Lagrangian can always be written in the form ${L = \tfrac{1}{2}G_{\alpha\beta}q'^{\alpha}q'^{\beta}+ V(q)}$ where $G_{\alpha\beta}$ is the metric tensor of the minisuperspace called the \textit{supermetric}, $V(q)$ is the \textit{superpotential}, and $q$ denotes the coordinates of the minisuperspace. From the reduced Lagrangian, the canonical momenta and Hamiltonian are calculated in the usual manner, we then promote the minisuperspace coordinates and canonical momenta to Hermitian operators and require that the Hamiltonian operator annihilates the physical states in the Hilbert space. The question is whether the symmetry-reduced theory yields the same field equations as the full theory when the spacetime metric is substituted in its field equations --- i.e., the variation commutes with the symmetry reduction. If such \textit{equivalence} holds for any theory, the symmetry is said to obey the \textit{principle of symmetric criticality (PSC)} \cite{Fels:2001rv,Anderson1997,Anderson:1999cn,Anderson:1999cm} (see also \cite{Palais:1979rca,Torre:2010xa,Frausto:2024egp}). This is a major limitation of minisuperspace quantization. It can lead to wrong field equations and a non-unique symmetry reduction (when PSC1 is violated) or to an insufficient number of field equations (when PSC2 is violated). The latter means extra spurious solutions, which may imply that, upon quantization, the Hilbert space is larger than the physical Hilbert space. In principle, the field equations of the reduced theory could coincide with the reduced field equations of the full theory only for specific theories (even if PSC is violated), or for metric ansatzes that go beyond those fully captured by symmetries. However, the first possibility cannot be analyzed systematically, since no restricted versions of PSC1 and PSC2 are known for subclasses of theories. Nevertheless, the equivalence can always be checked directly, and in GR it does not occur for PSC-violating minisuperspace reductions. The second possibility does not admit a rigorously defined Lagrangian reduction in the first place. For example, ad hoc ansatz reductions do not come with a canonical object playing the role of the $l$-chain. Consequently, in GR, PSC-compatible minisuperspace models seem to be the only ones admitting a consistent and rigorous symmetry reduction of the Lagrangian, and hence a well-defined canonical quantization.

Already at the classical level, the minisuperspace may possess symmetries, called \textit{conditional symmetries (CSs)}, which correspond to conformal Killing vectors of the supermetric that scale the superpotential consistently \cite{Christodoulakis:2001um,Paliathanasis:2024cbk,Dimakis:2016psv,Christodoulakis:2000xe,Christodoulakis:2018swq}. They arise in constrained systems as symmetries that leave the solutions of the full system invariant only when the primary constraints are applied; in other words, they do not act as symmetries of the system without the primary constraints, which is why they are called ‘conditional’ symmetries. Like spacetime symmetries, CSs lead to conserved quantities through the first Noether theorem. In minisuperspace quantization, CSs are used to define \textit{constants of motion (COMs)} commuting with the Hamiltonian operator, and the wave function must be an eigenfunction of them (once promoted to operators). This property is used to simplify the WDW equation. As we shall see, imposing all eigenvalue equations from the COMs often leads to a trivial wave function; thus, only subalgebras of the full algebra of COMs are applied, leaving room for non-trivial wave functions \cite{Christodoulakis:2012eg}.

A problem associated with this type of quantization is the \textit{nonuniqueness of the measure} in the minisuperspace. There have been attempts to select relevant measures in \cite{Kuchar:1991qf,Halliwell:1988wc}, but the problem remains open. We can impose conditions to fix the measure such as rescaling the supermetric and superpotential so that the superpotential is a constant, in this gauge the CS generators are the KVs of the supermetric. Other conditions include demanding a normalizable probability distribution, require the operators of the CS algebra or a selected subgroup of it to be Hermitian, or using the Newton-Wigner coordinates where the probability distribution would be the norm square of the wave function directly. However, all the methods mentioned are gauge-dependent and till now there is no concrete way to fix the measure on minisuperspaces, and it remains an open problem.

In this paper, we carry out canonical quantization of all PSC-compatible minisuperspace models, i.e., those admitting a consistent and rigorous symmetry reduction. We write the reduced Lagrangians, derive the algebras of CSs (and their subalgebras) and the corresponding COMs, and solve the WDW equation for both general wave functions and with imposed subalgebras of the COM algebra \cite{Christodoulakis:2012eg}. The paper is structured as follows:
\begin{itemize}
    \item In Sec.~\ref{The principle of symmetric criticality}, we review symmetry reduction of the Lagrangian and PSC, which provides the main criterion for rigorously reducible (and thus meaningfully quantizable) minisuperspace models, together with PSC-compatible symmetries $[d,l,c]$ within the Hicks classification \cite{Hicks:thesis,Frausto:2024egp}.
    \item In Sec.~\ref{Quantization of $[4,3,-]$}, we employ the minisuperspace quantization procedure to metrics with symmetries $[4,3,-]$. This includes symmetries of spherical, hyperbolic, and planar (s./h./p.) Schwarzschild and Taub--NUT spacetimes, BI/BII/BIII metrics, near-horizon extreme Kerr (NHEK) geometry, and the swirling universe, all of which are vacuum solutions of GR. We derive the CSs of their minisuperspace and find the wave functions of the corresponding quantum systems.
    \item In Sec.~\ref{Quantization of FLRW spaces ($[6,3,-]$)}, we apply the same procedure to quantize FLRW spacetimes ($[6,3,\{1 - 3\}]$) and their less-known counterparts ($[6,3,\{4 - 6\}]$). We first consider the vacuum FLRW ansatz, in which case the only classical vacuum solution is Minkowski spacetime (for flat or open FLRW), which is maximally symmetric rather than having the desired strictly FLRW symmetry. To address this, we also consider extensions with a cosmological constant and a massless scalar field. 
    \item In Sec.~\ref{Sec.[3,3,-]}, we perform the quantization of Bianchi type I and II spacetimes. Because Bianchi types VIII and IX are difficult to quantize, and no exact vacuum GR solutions that can be written in closed forms exist with these symmetries, we consider only special cases where adding certain Killing vectors reduces the model to a previously quantized class, namely $[4,3,-]$ or $[6,3,-]$.
    \item The paper is concluded in Sec.~\ref{Summary and conclusions}, which contains a summary of the results and possible directions for future research.
\end{itemize}

\section{Symmetry reduction of Lagrangians}\label{The principle of symmetric criticality}
PSC guarantees the equivalence of the field equations calculated by variation of the reduced theory with the reduced field equations (obtained from the full theory) simultaneously for all Lagrangians. As shown in \cite{Fels:2001rv} (see also \cite{Palais:1979rca,Anderson1997,Anderson:1999cn,Anderson:1999cm} for previous works), there exist two conditions, PSC1 and PSC2, on the spacetimes symmetries (given by the Lie algebras of Killing vector fields) that are satisfied if and only if PSC holds. All possible symmetries when PSC holds were identified in \cite{Frausto:2024egp} by direct checks of PSC1 and PSC2 with respect to the modern mathematical classification of infinitesimal group actions.

Let $M$ be a $4$-dimensional manifold and $\Gamma$ a $d$-dimensional Lie algebra of vector fields on $M$ generating an infinitesimal action of some Lie group. To distinguish in-equivalent actions (e.g., with the same abstract algebra) one needs to analyze their isotropy subalgebras $\Gamma_{\mathrm{x}}$, which are realized by those vector fields that leave the point ${\mathrm{x} \in M}$ unchanged, ${\Gamma_{\mathrm{x}} = \{\,X \in \Gamma \mid X|_{\mathrm{x}} = 0\,\}}$. Note that the quotient $\Gamma/\Gamma_{\mathrm{x}}$ is the tangent space to the orbit of $\mathrm{x}$. If ${\lie_X g=0}$, ${X\in\Gamma}$, then $X$ and ${g}$ are called the \textit{Killing vectors} and the \textit{symmetry-invariant metric}. The isotropy subalgebras for the Lorentzian metric correspond to the subalgebra of the Lorentz algebra. Whenever the spacetime locally splits into an isotropy-preserving slice and a group orbit, it can be classified by the abstract \textit{Lorentzian pairs} of algebra-subalgebra [corresponding to $(\Gamma,\Gamma_{\mathrm{x}})$], where the subalgebra acts on the quotient as a subalgebra of Lorentz algebra. We refer to this classification of spacetime symmetries as the \textit{Hicks classification} \cite{Hicks:thesis} (extension of \cite{Petrov,Fels_Renner_2006,Bowers:2012,Snobl2014-te,Rozum2015-lp}). The individual infinitesimal group actions in this classification are denoted by $[d,l,c]$, which respectively capture the dimension of ${\Gamma}$, the dimension of the orbit ${\Gamma}/{\Gamma}_{\mathrm{x}}$ (with ${p=d-l}$ being the dimension of the isotropy ${\Gamma}_{\mathrm{x}}$), and an extra distinguishing label. Since our interest lies in minisuperspace models, we focus on hypersurface-homogeneous spacetimes $[-,3,-]$ (i.e., ${l=3}$), where the space of orbits (the reduced space) is 1-dimensional (with 3-dimensional orbits), leading to ordinary differential equations upon reduction for the single-variable metric functions. The cases $[-,\{1,2\},-]$ correspond to midisuperspace models with 3- and 2-dimensional space of orbits resulting in partial differential equations. The cases $[-,4,-]$, on the other hand, describe the homogeneous spacetimes with 0-dimensional space of orbits and give rise to algebraic equations. Although the symmetry-reduction is still possible in $[-,4,-]$, where the reduced Lagrangian depends on constants describing the symmetry-invariant metrics, it does not seem to define a meaningful quantum theory.

In contrast to the reduction of the field equations $E(L)[g]$, which correspond to an ordinary substitution of the symmetry-invariant metric $E(L)[\check{g}]$, ${\lie_X \check{g}=0}$, the rigorous reduction of the Lagrangian $L[g]$, when understood as an anti-symmetric top-form (rather than as an integrable density) also involves the reduction of the form degree from 4 to the ${4-l}$, which we denote by $\check{L}[\check{g}]$. Instead of a problematic integration over the orbit, which may lead to diverging integrals, this can be rigorously accomplished by contraction with a symmetry-invariant anti-symmetric $(l,0)$-tensor $\chi$ called the \textit{$l$-chain}. In some basis of Killing vector fields $X_i$, ${i=1,\dots,d}$, it takes the form
\begin{equation}
    {\chi}=\sum_{\mathclap{i_1<\dots<i_l}}\chi^{i_1\dots i_l} {X}_{i_1}\wedge\cdots\wedge{X}_{i_l},
\end{equation}
where $\wedge$ denotes the exterior product in the contravariant indices and $\chi^{i_1\dots i_l}$ are its symmetry-invariant scalar components. The reduced Lagrangian is then canonically (up to an arbitrary scaling constant) given by
\begin{equation}\label{eq:redLagr}
    \check{L}[q_i]:=L[\check{g}]\bullet \chi, 
\end{equation}
where $\bullet$ stands for the contraction of the last indicies of $L$ with all indicies of $\chi$ and $q_i$ being the symmetry-invariant scalars given by the expansion of the symmetry-invariant metric in the symmetry-invariant symmetric ${(0,2)}$-tensor basis $\varsigma_i$, ${\check{g}=\sum_{i=1}^d q_i \varsigma_i}$. The coefficients $q_i$ are the scalar fields living on the reduced space, which play a role of the minisuperspace coordinates.

Having established the reduced Lagrangian, we can now formulate PSC as the equivalence between reduced field equations and field equations of the reduced theory for all Lagrangians,
\begin{equation}\label{eq:PSC}
    \forall {L}: {{E}}({L})[\check{{g}}]=0 \Longleftrightarrow {{E}}_i(\check{{L}})[q_j]=0,
\end{equation}
where ${{E}}({L})[\check{{g}}]$ is the Euler--Lagrange form of the full theory $L$ evaluated on the reduced metric $\check{{g}}$ while ${{E}}_i(\check{{L}})[q_j]$ are the Euler-Lagrange forms of the reduced theory $\check{{L}}$. The statement \eqref{eq:PSC} can be reformulated as two independent conditions (PSC1 and PSC2) for the infinitesimal group actions, for which we refer the reader to \cite{Fels:2001rv}; they were checked with respect to the Hicks classification in \cite{Frausto:2024egp}.\footnote{Remark that the infinitesimal group actions not classifiable by the Hicks classification do not satisfy PSC.} In what follows, we focus on all PSC-compatible symmetry-invariant metrics with 3-dimensional orbits, i.e., $[3,3,\{2,3,8,9\}]$, $[4,3,\{1-6,8-11\}]$, and $[6,3,\{1-6\}]$. It is worth mentioning that some of these metrics are related via double Wick rotations, namely, $[4,3,\{8-11\}]$ are double Wick rotations of $[4,3,\{1-6\}]$ as pointed out in \cite{Colleaux:2025uiw}. 
We will show that these double Wick rotations do not affect the quantization process in Sec.~\ref{Wick rotations}.

Note that symmetry-invariant metrics $\check{g}$ may possess a residual diffeomorphism freedom, which allows one to fix some of the functions $q_i$ to specific forms, as is often done in the literature. While such gauge-fixing is harmless at the level of the field equations (i.e., after variation), imposing it directly at the level of the reduced Lagrangian (before variation) can lead to the loss of important field equations, even when the PSC is satisfied \cite{Frausto:2024egp}. In some cases, equivalence can still hold despite the gauge-fixing, which may sometimes be justified by the Noether identities associated with the infinite-dimensional part of the residual diffeomorphism group. Nevertheless, questions naturally arise concerning the gauge-independence of the quantization procedure. To avoid such issues, we refrain from using the residual diffeomorphism freedom at any stage of our calculations.

\section{Symmetries of s./h./p. Schwarzschild/Taub--NUT and their double Wick rotations --- $[4,3,-]$}\label{Quantization of $[4,3,-]$}
In this section, we canonically quantize the symmetry-invariant metrics possessing the symmetries of s./h./p. Schwarzschild, s./h./p. Taub--NUT, and their double Wick rotations. For both Schwarzschild and Taub--NUT, the equations and discussion are analogous in the spherical and hyperbolic cases, thus, we group them together. The planar case is studied in a separate subsection, where we see differences in CSs and the quantization procedure. All these classes of metrics contain classical vacuum GR solutions.

\subsection{Spherical and hyperbolic Schwarzschild --- $[4,3,\{3,1\}]$}

\subsubsection{Symmetry-invariant metric and classical solutions}
The Killing vectors generating the symmetries of s./h. Schwarzschild are given by
\begin{equation}
    \begin{aligned}
        X_1 = \sqrt{1-k\rho}\left(\cos\varphi\partial_{\rho} - \tfrac{\sin\varphi}{\rho}\partial_{\varphi}\right), \quad X_2 = \sqrt{1-k\rho}\left(\sin\varphi\partial_{\rho} - \tfrac{\cos\varphi}{\rho}\partial_{\varphi}\right), \quad X_3 = \partial_{\varphi}, \quad X_4 = \partial_t,
    \end{aligned}
\end{equation}
where $k=1$ for spherical and $k=-1$ for hyperbolic cases. The symmetry-invariant metric, i.e., the general metric satisfying $\lie_{X_i} \check{g}=0$, can be written as 
\begin{equation}\label{inv metric1}
    \check{g} = - q_1(r) dt^2 + 2 q_2(r) dtdr + q_3(r)dr^2 + q_4(r)\left(\tfrac{d\rho^2}{1-k\rho^2}+\rho^2d\varphi^2\right).
\end{equation}
The only classical vacuum GR solution (up to gauge transformations) is $q_1(r)= \tfrac{1}{q_3(r)} = k-\tfrac{2m}{r}$, $q_2(r)=0$ and $q_4(r)=r^2$, which corresponds to the s./h. Schwarzschild spacetimes (also known as AI/AII-metrics). In what follows, we derive the reduced Einstein--Hilbert Lagrangian, the CSs, and quantize the corresponding minisuperspace.

\subsubsection{Reduced Lagrangian and generalized momenta}

To obtain the reduced Lagrangian of GR, we first need the Ricci scalar of the symmetry-invariant metric \eqref{inv metric1},
\begin{equation}\label{Ricci scalar of s./h. Schwarzschild}
\begin{aligned}
  R =& \tfrac{1}{2 q_4^2 \left(q_1 q_3+q_2^2\right)^2}\bigg(q_2^2 \big(q_1 \left(8 k q_3 q_4-4 q_4 q_4''+q_4'^2\right)-2 q_4 \big(q_4 q_1''+2 q_1' q_4'\big)\big)+q_1^2 \big(4 k q_3^2 q_4+2 q_4 q_3' q_4'\\&+q_3 \left(q_4'^2-4 q_4 q_4''\right)\big) +4 k q_2^4 q_4+2 q_2 q_4 q_2' \left(q_4 q_1'+2 q_1 q_4'\right)+q_3 q_4^2 q_1'^2+q_1 q_4 \left(q_1' \left(q_4 q_3'-2 q_3 q_4'\right)-2 q_3 q_4 q_1''\right)\bigg) .
\end{aligned}
\end{equation}
After multiplying by the Levi-Civita tensor 
\begin{equation}
    \varepsilon=\tfrac{\rho q_4}{\sqrt{1-k \rho ^2}} \sqrt{q_1 q_3+q_2^2}dt\wedge dr\wedge d\rho\wedge d\varphi,
\end{equation}
adding an exact top-form (total-derivative term) to cancel the second derivatives, and contracting with the appropriate $l$-chain, namely, 
\begin{equation}
    \chi \propto \tfrac{\sqrt{1-k\rho^2}}{\rho} \partial_{t} \wedge \partial_{\rho} \wedge \partial_{\varphi}
\end{equation}
through \eqref{eq:redLagr} (see \cite{Frausto:2024egp} for the list of $l$-chains in PSC-compatible cases), we arrive at the reduced Lagrangian 1-form
\begin{equation}\label{spherical schwarzschild}
   \check{L} = \text{const}\left(2k\sqrt{q_1 q_3+q_2^2} + \frac{q_1'q_4'+q_1(q_4')^2/2q_4}{\sqrt{q_1 q_3+q_2^2}}\right)\mathrm{d}r.
\end{equation}
In what follows, and in all analogous calculations, we omit $\mathrm{d}r$ and the overall constant, and simply use $\check{L}$ for the remaining function of $r$ alone. The generalized momenta are given by
\begin{equation}\label{generalized momenta for schw.}
    p_1 = \tfrac{\partial \check{L}}{\partial q_1'} = \tfrac{q_4'}{\sqrt{q_1 q_3+q_2^2}}, 
    \quad 
    p_4 = \tfrac{\partial \check{L}}{\partial q'_4}= \tfrac{q'_1+ q_1 q_4'/q_4}{\sqrt{q_1 q_3+q_2^2}},
    \quad p_2=p_3=0,
\end{equation}
while the Hamiltonian reads 
\begin{equation}\label{hamiltonian for schwarzschild}
    H = \sqrt{q_1 q_3+q_2^2} \left(p_1p_4- \tfrac{q_1}{2q_4}p_1^2-2k\right).
\end{equation}

The supermetric and the superpotential can be identified from the reduced Lagrangian \eqref{spherical schwarzschild} by recasting it to the form 
\begin{equation}\label{eq:L}
    \check{L} = \tfrac{1}{2}\tilde{G}_{\alpha\beta}(q)q'^{\alpha}q'^{\beta}- \tilde{V}(q),
\end{equation}
where $q \in \{q_1,q_2,q_3,q_4\}$ and $\alpha,\beta \in \{1,2,3,4\}$. In this case, the supermetric and superpotential are given by
\begin{equation}\label{eq:Gtilde}
    \tilde{G}_{\alpha\beta} = \frac{1}{\sqrt{q_1 q_3+q_2^2}}\begin{pmatrix}
        0& 1&0&0\\1&\frac{q_1}{q_4}&0&0\\0&0&0&0\\0&0&0&0
    \end{pmatrix}, \quad \tilde{V} =- 2k\sqrt{q_1 q_3+q_2^2}.
\end{equation}
Notice that we rearranged the coordinates to $\{q_1,q_4,q_2,q_3\}$ for convenience since $q_2$ and $q_3$ will be dropped when reducing the minisuperspace as we will see. From this point onward, we shall consider the supermetric and superpotential as tensors on the minisuperspace \cite{Schmidt:2001wt}, which is covered by the coordinates $q_i$, $i\in \{1,2,3,4\}$. 

It turns out that the supermetric and the superpotential can be taken more generally as
\begin{equation}
    \check{L}=\tfrac{\tilde{N}}{2 }G_{\alpha\beta}(q)q'^{\alpha}q'^{\beta}-\tfrac{V(q)}{\tilde{N}},
\end{equation}
where $\tilde{N} = \tilde{N}(q)$ is an arbitrary function --- this is equivalent to \eqref{eq:L} in the sense that both lead to the same CSs and WDW equation \cite{Christodoulakis:2012eg}. Here, we will choose $\tilde{N}=1/{\sqrt{q_1q_3+q_2^2}}$ with the rescaled supermetric and superpotentials being
\begin{equation}
    G_{\alpha\beta} = 
    \begin{pmatrix}
        0&1&0&0\\1&\tfrac{q_1}{q_4}&0&0\\0&0&0&0\\0&0&0&0
    \end{pmatrix}, 
    \quad 
    V =- 2k.
\end{equation}
The rescaled Hamiltonian takes the form of \eqref{hamiltonian for schwarzschild} but without the square-root prefactor. Notice that the supermetric is degenerate, which is unsuitable for quantization. Thus, we restrict ourselves to a subspace with constant $q_2$ and $q_3$, where the induced supermetric is non-degenerate, this is possible since the conjugate momenta $p_2$ and $p_3$ are zero. We can then write the induced supermetric and superpotential as
\begin{equation}\label{supermetric and superpotential of s./h. Schwarzschild}
    G_{\alpha\beta} = 
    \begin{pmatrix}
        0&1\\1&\tfrac{q_1}{q_4}
    \end{pmatrix}, 
    \quad 
    V =- 2k.
\end{equation}
In other words, from now on, we denote $q \in \{q_1,q_4\}$ and $\alpha,\beta \in \{1,4\}$.
For the sake of eliminating repetition, in the rest of this paper, we only use the non-degenerate supermetric on a subspace of the minisuperspace without referring to the full degenerate one.

\subsubsection{Conditional symmetries}

CSs are classical symmetries of constrained systems such that they leave symmetry-invariant the solutions of the full system only if the primary constraints are applied \cite{Christodoulakis:2012eg}. The generators of the CSs are the conformal Killing vectors $\xi^{\alpha} = (\xi^1(q),\xi^4(q))$ of the supermetric\footnote{The CS generators of the full degenerate minisuperspace (the Killing vectors of the supermetric when rescaling $V$ to be constant) are those of the reduced minisuperspace plus additional generators: $\xi_{q_2}= f_1(q)\partial_{q_2},\xi_{q_3}= f_2(q)\partial_{q_3}$ where $f_1$ and $f_2$ are arbitrary functions, i.e., one extra family of generators for every degenerate direction.} that scale the superpotential consistently,
\begin{equation}\label{CSs2}
    \begin{aligned}
  \pounds_{\xi}G^{\alpha\beta} = \Phi G^{\alpha\beta}, \quad  \pounds_{\xi}V = \Phi V,
    \end{aligned}
\end{equation}
where $\Phi = \Phi(q)$. For the symmetry-invariant metrics having the symmetries of s./h. Schwarzschild ($[4,3,\{3,1\}]$), the solutions are
\begin{equation}
    \xi_1 = \left(-q_1,q_4\right), \quad \xi_2=\left(\frac{1}{\sqrt{q_4}},0\right), \quad \xi_3=\left(-\frac{q_1}{2\sqrt{q_4}},\sqrt{q_4}\right).
\end{equation}
The COMs on the superspace $Q_i=\xi^{\alpha}_ip_{\alpha}$ read
\begin{equation}
    Q_1 = -q_1p_1+q_4p_4, \quad Q_2 = \frac{p_1}{\sqrt{q_4}}, \quad Q_3 = -\frac{q_1 p_1}{2\sqrt{q_4}}+\sqrt{q_4}\, p_4,
\end{equation}
satisfying the Poisson brackets
\begin{equation}
    \begin{aligned}
        \{Q_1,H\}=0, \quad \{Q_2,H\}=0, \quad \{Q_3,H\}=0.
    \end{aligned}
\end{equation}
The Poisson brackets of the COMs are given by
\begin{equation}
    \{Q_1,Q_3\}=Q_3, \quad \{Q_2,Q_1\}= Q_2, \quad \{Q_2,Q_3\}=0.
\end{equation}
Thus, the generators form a closed algebra of Bianchi type VI with $h=-1$ in the notation of \cite{caroca2013bianchi}.

\subsubsection{Quantization}
Here, we quantize the system on the $2d$ reduced minisuperspace whose coordinates are $q_1$ and $q_4$. We promote the variables $q_1,q_4,p_1,p_4$ and $Q_i$ into operators as follows: $q_1 \to \hat{q}_1$ such that $\hat{q}_1\ket{\Psi}=q_1\ket{\Psi}$, where $\Psi$ is the wave function, and similarly for $q_4$.\footnote{Here, $\Psi$ is only a function of $q_1$ and $q_4$ because $p_2=p_3=0$ which, upon quantization, lead to the primary constraints on the wave function given by $\partial_{q_2}\Psi=\partial_{q_3}\Psi=0$, i.e, the wave function on the whole minisuperspace reduces to a function on the reduced one.} $p_1$ is promoted into $\hat{p}_{1}$ such that $\hat{p}_{1}\ket{\Psi}=-i\partial_{q_1}\ket{\Psi}$, and similarly for $p_4$. The classical COMs $Q_i$ are promoted to Hermitian operators by first writing them as \cite{Christodoulakis:1990tua}
\begin{equation}
    Q_i = \frac{1}{2\mu}\left(\mu Q_i+Q_i \mu\right),
\end{equation}
where $\mu$ is the measure on the minisuperspace. As mentioned in the introduction, the measure is not uniquely defined; in fact, it can be written as $\mu = e^{\lambda}\sqrt{-\det(G_{\alpha\beta})}$, where $\lambda=\lambda(q)$ is an arbitrary function. Promoting $Q_i$ to an operator $\hat{Q}_i$ yields
\begin{equation}\label{Herm.}
    \hat{Q}_i = -\frac{i}{2\mu}\left(\mu \xi_i^{\alpha}\partial_{\alpha} + \partial_{\alpha}(\xi_i^{\alpha}\mu) \right).
\end{equation}
For the operators $\hat{Q}_i$ to generate symmetries, they must act as a derivative on the wave function \cite{Christodoulakis:2012eg}, thus, we require the second term in \eqref{Herm.} to vanish. 

The eigenvalue equations arising from the COMs are $\hat{Q}_i\ket{\Psi} = \kappa_i\ket{\Psi}$, where $\kappa_i$ are constants. The secondary constraint of the system is the Hamiltonian constraint (the WDW equation). Although the Hamiltonian is given by
\begin{equation}\label{eq.1}
    H = \tfrac{1}{2}G^{\alpha\beta}p_{\alpha}p_{\beta}+V,
\end{equation}
a naive promotion into operators leads to a non-Hermitian operator; also,  there are ambiguities in the kinetic term since operators generally do not commute. To solve these problems, we follow \cite{Christodoulakis:1990tua} (which was used in \cite{Christodoulakis:2013sya} and \cite{Dimakis:2017qcf}) adopting the form of the Hamiltonian constraint, which is Hermitian as an operator and recovers \eqref{eq.1} classically: 
\begin{equation}\label{hamiltonian constraint}
   \left( \tfrac{-1}{2\mu}\partial_{\alpha}\left[\mu G^{\alpha\beta}\partial_{\beta}\right]+V\right)\Psi=0,
\end{equation}
where $\Psi=\Psi(q)$ is the wave function of the quantized system. To quantize the geometry, we solve the above Hamiltonian constraint as well as the eigenvalue equations of the CSs. Substituting with the supermetric and superpotential, the WDW equation reads
\begin{equation}
    q_1^2 \partial_{q_1}^2\Psi-2q_1 q_4 \partial_{q_1}\partial_{q_4}\Psi+ q_1\partial_{q_1}\Psi- 4k q_1 q_4\Psi=0.
\end{equation}
The solution to this equation is given by
\begin{equation}\label{WF for [4,3,3]}
    \Psi_z = q_1^{-i z/2}\left(A J_{i z}\left(4\sqrt{k q_1q_4}\right)+B Y_{i z}\left(4 \sqrt{k q_1q_4}\right)\right),
\end{equation}
where $J_n(x)$ is the Bessel function of the first kind, $Y_n(x)$ is the Bessel function of the second kind, $z$ is a complex separation constant and $A$ and $B$ are complex normalization constants.
Notice that renaming $q_1=a^2$, $q_4=b^2$ and setting $k=1$ recover the results from \cite{Christodoulakis:2012eg}. Notice that when $q_1,q_4 \to \infty$ the absolute value of the wave function for s. Schwarzschild goes to zero, but that of h. Schwarzschild goes to zero only when $A \neq 0$ but diverges if $A=0$.

The above wave function is derived only by solving \eqref{hamiltonian constraint} while disregarding the eigenvalue equations for the COMs: $\hat{Q}_i\ket{\Psi} = \kappa_i\ket{\Psi}$.
Substituting with the forms for the COMs, we get a system of partial differential equations consisting of the WDW equation and three eigenvalue equations:
\begin{equation}
    \begin{gathered}
         q_1^2 \partial_{q_1}^2\Psi-2q_1 q_4 \partial_{q_1}\partial_{q_4}\Psi+ q_1\partial_{q_1}\Psi- 4k q_1 q_4\Psi=0,
        \\
        -i\left(-q_1 \partial_{q_1}\Psi +q_4 \partial_{q_4}\Psi\right)= \kappa_1 \Psi, 
        \quad
        -i\frac{1}{\sqrt{q_4}} \partial_{q_1} \Psi=\kappa_2 \Psi, 
        \quad
        -i\left(\sqrt{q_4}\partial_{q_4}\Psi-\frac{q_1}{2\sqrt{q_4}}\partial_{q_1}\Psi\right)=\kappa_3\Psi,
    \end{gathered}
\end{equation}
which has no solution except the trivial one $\Psi=0$. For this reason, we restrict ourselves to subalgebras of the CS algebra.  
Using the notation $\langle Q_1,Q_2,\cdots \rangle$ to refer to the subalgebra generated by $\{Q_1,Q_2, \cdots\}$, we find that the two non-Abelian subalgebras $\langle Q_1,Q_2\rangle$ and $\langle Q_2,Q_3\rangle$ result in trivial wave functions.\footnote{This can be considered a generalization of the results in \cite{Christodoulakis:2012eg} where the symmetry-invariant metric was compared to the Schwarzschild metric, and $\kappa_1$, $\kappa_2$, and $\kappa_3$ were derived to be proportional to the mass, the speed of light and the reciprocal to the speed of light respectively. Since considering the full algebra of CSs, or its two $2d$ non-Abelian subalgebras, forces $\kappa_2=\kappa_3=0$, the mentioned algebras were deemed to be inconsistent and were discarded. In this paper, we obtained the same end result from quantizing the full class of metrics \eqref{inv metric1} without giving $\kappa_1,\kappa_2$ or $\kappa_3$ any interpretation showing that the full algebra and the two $2d$ non-Abelian subalgebras have zero wave functions.} Imposing the conditions of the $2d$ Abelian subalgebra $\langle Q_2,Q_3\rangle$ or the $1d$ subalgebra $\langle Q_2 \rangle$ leads to the wave function 
\begin{equation}\label{wave function of <Q_2,Q_3>}
    \Psi = A e^{\tfrac{i}{2}\left(2 \kappa_2 q_1 \sqrt{q_4}+ \tfrac{8 k \sqrt{q_4}}{\kappa_2}\right)},
\end{equation}
    where $A$ is a complex normalization constant. Similarly, restricting to $\langle Q_3 \rangle$ leads to the same wave function but with the reparameterization $\kappa_3 = \tfrac{2k}{\kappa_2}$.
     Finally, the subalgebra $\langle Q_1 \rangle$ yields the wave function 
\begin{equation}\label{wave function of <Q_1>}
    \Psi(q) = q_1^{-i\kappa_1/2}\left(A J_{i \kappa_1}\left(4 \sqrt{k q_1q_4} \right)+B Y_{i \kappa_1}\left(4\sqrt{kq_1 q_4}\right)\right),
\end{equation}
where $A$ and $B$ are arbitrary complex constants. 

\subsection{Planar Schwarzschild --- $[4,3,6]$}

\subsubsection{Symmetry-invariant metric and classical solutions}
The Killing vectors defining the symmetries of p. Schwarzschild are
\begin{equation}
    X_1 = \cos\varphi \partial_{\rho} - \tfrac{\sin\varphi}{\rho}\partial_{\varphi},\quad X_2 = \sin\varphi\partial_{\rho}+\tfrac{\cos\varphi}{\rho}\partial_{\varphi},\quad X_3 = \partial_{\varphi},\quad X_4 = \partial_t.
\end{equation}
The general symmetry-invariant metric can be written as
\begin{equation}\label{inv metric2}
    \check{g} = -q_1(r)dt^2 + 2 q_2(r)dtdr + q_3(r)dr^2+ q_4(r)\left(d\rho^2+\rho^2d\varphi^2\right) .
\end{equation}
Within this metric ansatz, there exists a classical vacuum solution of GR (up to gauge transformations) given by $q_1(r)= \tfrac{1}{q_3(r)} = -\tfrac{2m}{r}$, $q_2(r)=0$ and $q_4(r)=r^2$, where $m$ can be gauge-fixed to an arbitrary non-zero constant. 

\subsubsection{Reduced Lagrangian and generalized momenta}
To find the reduced GR Lagrangian, we first calculate the Ricci scalar of the symmetry-invariant metric \eqref{inv metric2}. It leads to \eqref{Ricci scalar of s./h. Schwarzschild} but with ${k=0}$; hence the reduced Lagrangian is given by \eqref{spherical schwarzschild} with ${k=0}$. Although the Ricci scalar and some other expressions differ only by the value of $k$ and can be computed for general $k$, this is not the case for other quantities, as will be seen in the calculation of the CSs.

Rewriting the reduced Lagrangian to the form 
\begin{equation}
    \check{L} = \tfrac{1}{2\sqrt{q_1 q_3+q_2^2}}G_{\alpha\beta}(q)q'^{\alpha}q'^{\beta}\\-\sqrt{q_1 q_3+q_2^2}V(q),
\end{equation} 
we obtain the same supermetric as \eqref{supermetric and superpotential of s./h. Schwarzschild} but with a vanishing superpotential.
The generalized momenta and the Hamiltonian are given by \eqref{generalized momenta for schw.} and \eqref{hamiltonian for schwarzschild}, respectively, but with ${k=0}$.

\subsubsection{Conditional symmetries}
Writing the generators of CSs as $\xi^{\alpha}= (\xi^1(q),\xi^4(q))$, we can rewrite the equations
 \begin{equation}
     \begin{aligned}
  \pounds_{\xi}G^{\alpha\beta} = \Phi G^{\alpha\beta}, \quad  \pounds_{\xi}V = \Phi V,
    \end{aligned}
 \end{equation}
where $\Phi = \Phi(q)$, as
\begin{equation}\label{equations}
\begin{aligned}
    &\partial_{q_1} \xi^4=0, \quad \partial_{q_1}\xi^1+\partial_{q_4}\xi^4=\Phi,\quad  q_4 \xi^1 - q_1 \xi^4 + 2 q_4^2 \partial_{q_4}\xi^1 + 2q_1 q_4 \partial_{q_4}\xi^4 = q_1 q_4 \Phi.
    \end{aligned}
\end{equation}
Differentiating the third equation twice with respect to $q_1$ and using the second equation, we get an equation for $\Phi$:
\begin{equation}
    q_1^2\partial_{q_1}^2\Phi - 2q_1q_4 \partial_{q_1}\partial_{q_4}\Phi + q_1 \partial_{q_1}\Phi=0.
\end{equation}
The solution is given by
\begin{equation}
    \Phi = F(q_1^2q_4)+ A \ln q_1,
\end{equation}
where $F$ is an arbitrary function and $A$ is a constant. Substituting in the first derivative of the third equation with respect to $q_1$ and simplifying, we get
\begin{equation}
    \left(2q_4^2+q_4\right)\xi^4{}'(q_4)-\xi^4(q_4)=-Aq_4.
\end{equation}
The general solution is
\begin{equation}\label{g function}
    \xi^4 = \frac{q_4(C - A \ln q_4)}{1+2q_4},
\end{equation}
where $C$ is a constant. Substituting into the second equation in \eqref{equations} and solving for $\xi^1(q)$, we get
\begin{equation}\label{f function}
    \xi^1 = \int F(q_1^2q_4)dq_1 + A \left(q_1 \ln q_1 - q_1\right) - q_1\tfrac{C - A \ln q_4 - A - 2Aq_4}{(1+2q_4)^2} + f_1(q_4),
\end{equation}
where $f_1(q_4)$ is an arbitrary function of $q_4$. The relation between $F(q_1^2q_4)$ and $f_1(q_4)$ is given by the third equation in \eqref{equations}:
\begin{equation}
    \begin{aligned}
        &\int F(q_1^2q_4)dq_1 + 2q_4 \int F'(q_1^2q_4)q_1^2dq_1 + f_1(q_4)+2q_4 f'_1(q_4)-\tfrac{q_1(C+A-A\ln q_4)}{1+2q_4} + \tfrac{2A - 4Aq_4+ q_1(C-A\ln q_4)}{(1+2q_4)^2} 
        \\&
        +\tfrac{8q_4(C-A \ln q_4)}{(1+2q_4)^3} = q_4 \left(F(q_1^2q_4)+A\right).
    \end{aligned}
\end{equation}

In summary, the generators of the CSs are given by $\xi^{\alpha} = (\xi^1(q),\xi^4(q_4))$ where $\xi^1$ and $\xi^4$ are given by \eqref{f function} and \eqref{g function}, respectively. There is a free function in the definition of $\xi^1$, therefore, the algebra is infinite-dimensional. This means that we can construct an infinite number of constants of motion of the forms
\begin{equation}
    Q(\xi^1,\xi^4)= \xi^1p_1+\xi^4p_4.
\end{equation}
The Poisson bracket of $Q(\xi^1,\xi^4)$ with the Hamiltonian is given by
\begin{equation}
    \{Q(\xi^1,\xi^4),H\}= p_1p_4 \left(\partial_{q_1}\xi^1+\partial_{q_4}\xi^4\right)+ \tfrac{p_1^2}{2q_4^2}\left(q_4 \xi^1 - q_1 \xi^4 + 2 q_4^2 \partial_{q_4}\xi^1 - 2q_1 q_4 \partial_{q_1}\xi^1\right).
\end{equation}
Using \eqref{equations}, we can write
\begin{equation}
    \{Q(\xi^1,\xi^4),H\} = p_1p_4\Phi+ \tfrac{p_1^2}{2q_4^2}\left( q_4 \xi^1 - q_1 \xi^4 + 2 q_4^2 \partial_{q_4}\xi^1 + 2q_1 q_4 \partial_{q_4}\xi^4 - 2q_1 q_4 \Phi\right) = \left(p_1p_4-\tfrac{q_1p_1^2}{2q_4}\right)\Phi = H \Phi.
\end{equation}
That is, the Poisson bracket is weakly vanishing, meaning that it vanishes on the surfaces where the Hamiltonian constraint ($H=0$) is satisfied.

\subsubsection{Quantization}

The system is quantized by promoting the coordinates and canonical momenta to operators as before. The WDW equation in this case is
\begin{equation}
    \partial_{\alpha}(\mu G^{\alpha\beta}\partial_{\beta})\Psi=0.
\end{equation}
Substituting the supermetric, we get
\begin{equation}
q_1^2 \partial_{q_1}^2\Psi-2q_1 q_4 \partial_{q_1}\partial_{q_4}\Psi+ q_1\partial_{q_1}\Psi=0,
\end{equation}
whose solution reads
\begin{equation}\label{WF for [4,3,6]}
    \Psi(q_1,q_4) = \Psi_1\left(q_1 \sqrt{q_4}\right)+\Psi_2(q_4),
\end{equation}
for any functions $\Psi_1$ and $\Psi_2$. Thus, we have an infinite number of possible wave functions. We can simplify the WDW equation using CSs as in the previous sections. In this case, upon promoting $Q(\xi^1,\xi^4)$ to operators, we have an infinite number of eigenvalue equations of the form $\hat{Q}(\xi^1,\xi^4)\ket{\Psi} = \kappa(\xi^1,\xi^4)\ket{\Psi}$ where $\kappa(\xi^1,\xi^4)$ are constants with respect to $q$. Promoting the minisuperspace coordinates and momenta to operators, the eigenvalue equation can be written as
\begin{equation}
    \xi^1\partial_{q_1}\Psi+ \xi^4\partial_{q_4}\Psi = i \kappa(\xi^1,\xi^4) \Psi,
\end{equation}
for any functions $\xi^1$ and $\xi^4$ of the forms \eqref{f function} and \eqref{g function}. Since the equation must hold for any $\xi^1$ and $\xi^4$, we can set both to zero (can be done by setting all the arbitrary constants and functions in \eqref{f function} and \eqref{g function} to zero). On doing so, we get $\kappa(\xi^1,\xi^4)\Psi=0$. Thus, the solution is either $\Psi=0$ or $\kappa(\xi^1,\xi^4)=0$. In the former case, we have a zero wave function so we neglect this possibility, in the latter case the equation reads
\begin{equation}
    \xi^1\partial_{q_1}\Psi+ \xi^4\partial_{q_4}\Psi=0.
\end{equation}
Setting $\xi^1=0$ and $\xi^4 \neq 0$, we get $\Psi=\Psi(q_4)$, while setting $\xi^4=0$ and $\xi^1 \neq 0$ result in $\Psi=\Psi(q_1)$, this means that the only solution is $\Psi = \text{constant}$.

\subsection{Spherical and hyperbolic Taub--NUT --- $[4,3,\{4,2\}]$}\label{sec1}
\subsubsection{Symmetry-invariant metric and classical solutions}
The Killing vectors representing the symmetries of s./h. Taub--NUT are given by
\begin{equation}
    \begin{aligned}
        X_1 &= \sqrt{1-k\rho}\left(\cos\varphi\partial_{\rho} - \tfrac{\sin\varphi}{\rho}\partial_{\varphi}\right)-2n \tfrac{1-\sqrt{1-k\rho^2}}{k}\tfrac{\sin\varphi}{\rho}\partial_t, \\ X_2 &= \sqrt{1-k\rho}\left(\sin\varphi\partial_{\rho} - \tfrac{\cos\varphi}{\rho}\partial_{\varphi}\right)+2n \tfrac{1-\sqrt{1-k\rho^2}}{k}\tfrac{\cos\varphi}{\rho}\partial_t, \quad X_3 = \partial_{\varphi}, \quad X_4 = \partial_t,
    \end{aligned}
\end{equation}
where $k=1$ for s. Taub--NUT and $k=-1$ for h. Taub--NUT and $n$ is the NUT parameter. The corresponding symmetry-invariant metrics are
\begin{equation}\label{metric ansatz}
    \check{g} = - q_1(r)\left(dt+ 2n \tfrac{1-\sqrt{1-k\rho^2}}{k}d\varphi\right)^2 + 2 q_2(r)\left(dt +  2n \tfrac{1-\sqrt{1-k\rho^2}}{k}d\varphi \right)dr+ q_3(r)dr^2 + q_4(r)\left(\tfrac{d\rho^2}{1-k\rho^2}+\rho^2d\varphi^2\right) .
\end{equation}
For this class of metrics, there is a classical solution, up to gauge transformations, that corresponds to $q_1(r) = \tfrac{1}{q_3(r)}=\tfrac{k(r^2-n^2)-2mr}{r^2+n^2}$, $q_2(r)=0$ and $q_4(r)=r^2+n^2$.
In the next subsections, we find the reduced Einstein--Hilbert Lagrangian, calculate its CSs, and quantize the minisuperspace of this class of metrics.

\subsubsection{Reduced Lagrangian and generalized momenta}
To find the reduced GR Lagrangian, we calculate the Ricci scalar for the metric ansatz \eqref{metric ansatz},
\begin{equation}\label{ricci scalar}
\begin{aligned}
    R =& \tfrac{1}{2 q_4^2 \left(q_1 q_3+q_2^2\right)^2}\bigg(q_1 \big(q_2^2 \left(8 k q_3 q_4-4 q_4 q_4''+q_4'^2\right)+4n^2 q_2^4 +q_4 \left(q_1' \left(q_4 q_3'-2 q_3 q_4'\right)-2 q_3 q_4 q_1''\right)+4 q_2 q_4 q_2' q_4'\big)\\&+q_1^2 \left(4 k q_3^2 q_4+8 n^2 q_2^2 q_3+2 q_4 q_3' q_4'+q_3 \left(q_4'^2-4 q_4 q_4''\right)\right)+q_4 \big(4 k q_2^4+2 q_2 q_4 q_1' q_2'+q_3 q_4 q_1'^2-2 q_2^2 \big(q_4 q_1''\\&+2 q_1' q_4'\big)\big)+4n^2 q_1^3 q_3^2\bigg).
\end{aligned}
\end{equation}
The reduced Lagrangian derived from the symmetry reduction of $\varepsilon R$ is given by
\begin{equation}\label{general reduced lagrangian}
    \check{L} = \tfrac{1}{\sqrt{q_2^2+q_1q_3}}\left(q_1'q_4'+ \tfrac{q_1q_4'^2}{2q_4}\right) + \sqrt{q_2^2+q_1q_3}\left(2k +2n^2\tfrac{q_1}{q_4}\right),
\end{equation}
which we rewrite as
\begin{equation}\label{eq:checkLTNUT}
    \check{L} = \tfrac{1}{2\sqrt{q_2^2+q_1q_3}}G_{\alpha\beta}(q)q'^{\alpha}q'^{\beta}-\sqrt{q_2^2+q_1q_3}V(q),
\end{equation}
where $q=\{q_1,q_4\}$ are the generalized coordinates. On the minisuperspace, we can identify the supermetric and the superpotential,
\begin{equation}\label{eq:supmetpot}
    G_{\alpha\beta}= 
    \begin{pmatrix}
        0&1\\1&\tfrac{q_1}{q_4} 
    \end{pmatrix}, \quad V = -2k-2n^2\tfrac{q_1}{q_4}.
\end{equation} 
The generalized momenta are given by
\begin{equation}\label{generalized momenta for TNUT}
    p_1=\tfrac{q_4'}{\sqrt{q_1q_3+q_2^2}}, \quad p_4 = \tfrac{q_1'+q_1q_4'/q_4}{\sqrt{q_1q_3+q_2^2}}, \quad p_2=p_3=0,
\end{equation}
which can be used to express $q_1'$ and $q_4'$ as
\begin{equation}
    q_1'= \sqrt{q_1q_3+q_2^2}\, \, p_1, \quad q_4' = \sqrt{q_1q_3+q_2^2}\left(p_4-\tfrac{q_1}{q_4}p_1\right).
\end{equation}
The Hamiltonian is then obtained in the usual way; the result is
\begin{equation}\label{Hamiltonian for TNUT}
    H = \sqrt{q_1q_3+q_2^2} \left(p_1p_4 - \tfrac{q_1}{2q_4}p_1^2 - 2k - 2n^2 \tfrac{q_1}{q_4}\right).
\end{equation}
Notice that on the reduced minisuperspace where $q_2$ and $q_3$ are held constant, we have only two generalized momenta $p_1$ and $p_4$; the Hamiltonian to be considered is the same as \eqref{Hamiltonian for TNUT} without the prefactor $\sqrt{q_1q_3+q_2^2}$.

\subsubsection{Conditional symmetries}
 
Now, we derive the algebra of CSs. Their generators are given by the vector field $\xi^{\alpha} = (\xi^1(q),\xi^4(q))$ satisfying
\begin{equation}
    \begin{aligned}
  \pounds_{\xi}G^{\alpha\beta} = \Phi G^{\alpha\beta}, \quad  \pounds_{\xi}V = \Phi V.
    \end{aligned}
\end{equation}
The second equation gives the function $\Phi$ in terms of $\xi^1$ and $\xi^4$:
\begin{equation}
    \Phi =\tfrac{n^2}{q_4}\tfrac{q_4 \xi^1- q_1 \xi^4}{kq_4+ n^2 q_1}.
\end{equation}
Substituting into the first equation, we get 
\begin{equation}\label{CSs}
    \begin{aligned}
       &\partial_{q_1} \xi^4=0, \quad \partial_{q_1}\xi^1+\partial_{q_4}\xi^4=-\tfrac{n^2}{q_4}\tfrac{q_4 \xi^1- q_1 \xi^4}{kq_4+ n^2 q_1},\quad  q_4 \xi^1 - q_1 \xi^4 + 2 q_4^2 \partial_{q_4}\xi^1 - 2q_1 q_4 \partial_{q_4}\xi^1 = n^2 q_1 \tfrac{q_4 \xi^1- q_1 \xi^4}{kq_4+ n^2 q_1}.
    \end{aligned}
\end{equation}
From the first equation, we can write $\xi^4=\xi^4(q_4)$. The second and third equations give
\begin{equation}\label{extra eq}
    q_4 \xi^1-q_1\xi^4+2q_4^2\partial_{q_4}\xi^1- q_1q_4\partial_{q_1}\xi^1+q_1 q_4 \xi^4{}'=0.
\end{equation}
Dividing by $q_1$ and differentiating with respect to $q_1$, we get an equation for $\xi^1$,
\begin{equation}
    q_1\partial_{q_1}\xi^1 -\xi^1+ 2q_1q_4\partial_{q_1}\partial_{q_4}\xi^1-2q_4\partial_{q_4}\xi^1-q_1^2\partial_{q_1}^2\xi^1=0.
\end{equation}
The general solution of this equation is
\begin{equation}
    \xi^1 = q_1 F(q_4)+\tfrac{1}{\sqrt{q_4}}G\left(q_1\sqrt{q_4}\right),
\end{equation}
where $F$ and $G$ are arbitrary functions. We first deduce the form of $G$. To do that, we begin with the second equation in \eqref{CSs} and differentiate it twice with respect to $q_1$; the resulting equation reads
\begin{equation}\label{eq3}
    (kq_4 + n^2 q_1)\sqrt{q_4}G'''\left(q_1\sqrt{q_4}\right)= -3n^2 G''\left(q_1\sqrt{q_4}\right).
\end{equation}
The general solution for $G$ is given by
\begin{equation}
G\left(q_1\sqrt{q_4}\right) = C_1 + C_2 q_1\sqrt{q_4},
\end{equation}
where $C_1$ and $C_2$ are arbitrary constants. Now, substituting this solution into the first derivative of \eqref{extra eq} with respect to $q_1$, we obtain 
\begin{equation}
    q_4 \partial_{q_4}\xi^4= - \xi^4.
\end{equation}
The general solution is
\begin{equation}
    \xi^4 = \tfrac{A}{q_4},
\end{equation}
where $A$ is an arbitrary constant. The function $\xi^1$ can be written as
\begin{equation}
    \xi^1 = q_1F(q_4)+ \tfrac{C_1}{q_1} + C_2 \sqrt{q_4}.
\end{equation}
Using this form in the second equation of \eqref{CSs}, we get an algebraic equation in $F$:
\begin{equation}\label{g' relation}
   \left(\tfrac{kq_4}{n^2}+q_1\right)\left(q_4^2F(q_4)-C_1-A\right)=-q_1q_4^2 F(q_4)-\tfrac{C_1q_4^2}{q_1}-C_2q_4^2\sqrt{q_4}+Aq_1.
\end{equation}
Since $F$ is only a function of $q_4$, all orders of $q_1$ must vanish. This gives a system of algebraic equations:
\begin{equation}
    C_1=0, \quad q_4^2 F(q_4)=A,\quad \tfrac{4kA}{n^2}=-C_2q_4\sqrt{q_4}.
\end{equation}
The third equation implies that $C_2=A=0$, and the second equation requires that $F(q_4)=0$. Thus, the only solution for the system \eqref{CSs} is
\begin{equation}
    \xi^1 = \xi^4=0,
\end{equation}
meaning that there are no CSs.

\subsubsection{Quantization}
To quantize the system, we promote the generalized coordinates and momenta to operators as in the previous sections.
Since there are no CSs, the wave function is determined only by the solution of the WDW equation. Substituting our supermetric and superpotential into \eqref{hamiltonian constraint}, after some algebra, we end up with the WDW equation of the form
\begin{equation}\label{general taub nut eq.}
    q_1\partial_{q_1}\Psi + q_1^2\partial_{q_1}^2\Psi - 2q_1q_4 \partial_{q_1}\partial_{q_4} - 4kq_1q_4 - 4n^2 q_1^2=0,
\end{equation}
whose general solution is
\begin{equation}\label{WF for [4,3,(4,2)]}
     \Psi_z(q) = \left(A J_z\left(4 \sqrt{k q_1 q_4}\right)+B Y_z\left(4 \sqrt{k q_1q_4}\right)\right) \left(C I_{\frac{z}{2}}\left(2q_1 n\right)+D K_{\frac{z}{2}}\left(2q_1 n\right)\right),
\end{equation}
where $A,B,C$ and $D$ are complex constants. To see the asymptotic behavior of the wave function as $q_1,q_4 \to \infty$, we use the asymptotic properties of the Bessel functions: $J_z(x) \sim \tfrac{1}{\sqrt{x}}\cos\left(x-\tfrac{\pi z}{2}-\tfrac{\pi}{4}\right)$, $Y_z(x) \sim \tfrac{1}{\sqrt{x}}\sin\left(x-\tfrac{\pi z}{2}-\tfrac{\pi}{4}\right)$, $I_z(x) \sim \tfrac{e^{x}}{\sqrt{x}}$ and $K_z(x)\sim \tfrac{e^{-x}}{\sqrt{x}}$. For $k>0$, when $C \neq 0$ the absolute value of the wave function for Taub--NUT is unbounded at infinity ($q_1,q_4 \to \infty$).
When $k<0$, we use the identities $J_z\left(4i\sqrt{\abs{k}q_1q_4}\right)=i^z I_z\left(4\sqrt{\abs{k}q_1q_4}\right)$ and $Y_z\left(4i\sqrt{\abs{k}q_1q_4}\right)=i^z K_z\left(4\sqrt{\abs{k}q_1q_4}\right)$ to derive the asymptotic behavior of the wave function. When $C\neq 0$, $\abs{\Psi_z}\to \infty$, but when $C=0$ the limit of the absolute value of the wave function is governed by the quantity $F=4\sqrt{\abs{k}q_1q_4}-2nq_1$: if $F \to \infty$ when $q_1,q_4 \to \infty$ then $\abs{\Psi_z} \to \infty$, otherwise $\abs{\Psi_z}\to 0$.

\subsection{Planar Taub--NUT --- $[4,3,5]$}

\subsubsection{Symmetry-invariant metric and classical solutions}
The Killing vectors of the p. Taub--NUT are
\begin{equation}
    X_1 = \cos\varphi\partial_{\rho} - \tfrac{\sin\varphi}{\rho}\partial_{\varphi}-\tfrac{2n \rho \sin\varphi}{2}\partial_t,\quad X_2 = \sin\varphi\partial_{\rho}+\tfrac{\cos\varphi}{\rho}\partial_{\varphi}+ \tfrac{2n\rho \cos\varphi}{2}\partial_t,\quad X_3 = \partial_{\varphi},\quad X_4 = \partial_t.
\end{equation}
The general symmetry-invariant metric reads
\begin{equation}\label{Inv metric for p.TNUT}
    \check{g} = -q_1(r)\left(dt+n\rho^2d\varphi\right)^2 + 2 q_2(r)\left(dt+ n\rho^2d\varphi\right)dr + q_3(r)dr^2+ q_4(r)\left(d\rho^2+\rho^2d\varphi^2\right).
\end{equation}
For this class of metrics, there is a classical solution for GR (up to gauge transformations) corresponding to $q_1(r) = \tfrac{1}{q_3(r)} = \tfrac{-2mr}{r^2+n^2}$, $q_2(r)=0$ and $q_4(r)=r^2+n^2$.

\subsubsection{Reduced Lagrangian and generalized momenta}

The formula for the Ricci scalar of the symmetry-invariant metric is given by \eqref{ricci scalar} but with $k=0$. We again emphasize that all quantities should be derived from the symmetry-invariant metric \eqref{Inv metric for p.TNUT}, as not all formulas from the previous section can be directly extended to ${k=0}$.
Nevertheless, the reduced Lagrangian coincides with \eqref{general reduced lagrangian} where ${k=0}$.
Also, the reduced Lagrangian can be written in the form \eqref{eq:checkLTNUT}, where the supermetric and the superpotential are \eqref{eq:supmetpot} except here $k=0$.
Consequently, the canonical momenta and Hamiltonian have the forms of \eqref{generalized momenta for TNUT} and \eqref{Hamiltonian for TNUT}, respectively, but with $k=0$.
\subsubsection{Conditional symmetries}
Let us calculate the CSs generated by the vector fields $\xi^{\alpha}=(\xi^1(q),\xi^4(q))$ as solutions of 
\begin{equation}
     \begin{aligned}
  \pounds_{\xi}G^{\alpha\beta} = \Phi G^{\alpha\beta}, \quad  \pounds_{\xi}V = \Phi V,
    \end{aligned}
 \end{equation}
where $\Phi=\Phi(q)$.
The equations for $\xi^1$ and $\xi^4$ are given by
\begin{equation}\label{k=0 eq}
    \begin{aligned}
        &\partial_{q_1} \xi^4 = 0, \quad \partial_{q_1}\xi^1+\partial_{q_4}\xi^4 = -\tfrac{q_4 \xi^1 - q_1 \xi^4}{q_1q_4},\quad
q_4 \partial_{q_4}\xi^1-q_1\partial_{q_1}\xi^1=0.
    \end{aligned}
\end{equation}
The general solution for these equations is $\xi^1(q)=A q_1q_4 + \tfrac{B}{q_1q_4}$ and $\xi^4(q_4)=-2Bq_4^2+Cq_4$, where $A,B,C$ are arbitrary constants. The generators of CSs are then given by 
\begin{equation}
    \xi^{\alpha}_1=(0,q_4), \quad \xi^{\alpha}_2=\left(\tfrac{1}{q_1q_4},0\right), \quad \xi^{\alpha}_3=\left(q_1q_4,-2q_4^2\right).
\end{equation}
Thus, we obtain three COMs $Q_i=\xi^{\alpha}_ip_{\alpha}$ corresponding to
\begin{equation}
    Q_1=q_4 p_4, \quad Q_2=\tfrac{p_1}{q_1q_4}, \quad Q_3=q_1q_4 p_1-2q_4^2 p_4.
\end{equation}
These are indeed conserved, since their Poisson brackets with the Hamiltonian satisfy
\begin{equation}
    \{Q_1,H\} = -H, \quad  \{Q_2,H\} = \tfrac{1}{q_1^2q_4}H, \quad \{Q_3,H\} = 3q_4 H,
\end{equation}
meaning that they vanish weakly --- that is, they are zero on the surfaces where the Hamiltonian constraint holds.
Their Poisson brackets are given by
\begin{equation}
    \{Q_1,Q_2\} = Q_2, \quad \{Q_1,Q_3\} = -Q_3, \quad \{Q_2,Q_3\} = 0,
\end{equation}
which shows that it is again a closed algebra of Bianchi type VI with $h=-1$.

\subsubsection{Quantization}

To quantize the system, we promote $q_1,q_4,p_1,p_4$ to operators as in the previous sections. The WDW equation can be written as
\begin{equation}
     q_1 \partial_{q_1}\Psi + q_1^2\partial_{q_1}^2\Psi - 2 q_1q_4 \partial_{q_1}\partial_{q_4}\Psi - 4n^2 q_1^2\Psi=0.
\end{equation}
The solution of this equation is given by
\begin{equation}\label{WF for [4,3,5]}
    \Psi_z(q) = (q_1q_4)^{z/2} \big[A I_{\tfrac{z}{2}}\left(2nq_1 \right)+B K_{\tfrac{z}{2}}\left(2n q_1 \right)\big],
\end{equation}
where $A$ and $B$ are arbitrary complex constants.

Imposing the constraints due to CSs, we demand the wave function to be an eigenfunction of the COMs, i.e., $\hat{Q}_i\ket{\Psi}=\kappa_i\ket{\Psi}$, where $\kappa_i$ are constants. Explicitly, this can be written as
\begin{equation}
    \begin{aligned}
        -iq_4 \partial_{q_4} \Psi=\kappa_1\Psi, \quad -\tfrac{i}{q_1q_4}\partial_{q_1}\Psi=\kappa_2 \Psi, \quad  -iq_1q_4\partial_{q_1} \Psi+ 2i q_4^2\partial_{q_4}\Psi=\kappa_3 \Psi.
    \end{aligned}
\end{equation}
Using these equations, the Hamiltonian constraint takes the form
\begin{equation}
    2i q_1^2 q_4 \kappa_2 \Psi - q_1^4 q_4 \kappa_2^2\Psi +2q_1^2q_4 \kappa_1\kappa_2 \Psi -4n^2 q_1^2  \Psi=0,
\end{equation}
which has only the trivial solution $\Psi=0$ for non-zero $n$. Notice that when $n=0$, the WDW equation coincides with its p. Schwarzschild counterpart. However, each equation must be solved separately and the limit $n \to 0$ of the wave function \eqref{WF for [4,3,5]} does not simplify directly to the wave function of p. Schwarzschild.

As in the previous cases, the full algebra gives only the trivial solution, so we restrict ourselves to subalgebras. The non-Abelian subalgebras $\langle Q_1,Q_2 \rangle$ and $\langle Q_1,Q_3\rangle$ result in trivial wave functions, while the $2d$ Abelian subalgebra $\langle Q_2,Q_3\rangle$ as well as $\langle Q_2 \rangle$ leads to
\begin{equation}
    \Psi(q) = A e^{\tfrac{i\kappa_2 q_1^2q_4}{2}-\tfrac{2in^2}{\kappa_2 q_4}},
\end{equation}
where $A$ is an arbitrary complex constant. The subalgebra $\langle Q_3 \rangle$ produces the same wave function with the reparameterization $\kappa_2 = -\tfrac{4n^2}{\kappa_3}$. 
Finally, restricting to $\langle Q_1 \rangle$ gives the wave function
\begin{equation}
    \Psi(q)=(q_1q_4)^{i \kappa_1} \big[A I_{i \kappa_1} \left(2 n q_1\right)+B K_{i \kappa_1} \left(2 n q_1\right)\big],
\end{equation}
where $A$ and $B$ are arbitrary complex constants.

\subsection{Double Wick rotations --- $[4,3,\{8,9,10,11\}]$}\label{Wick rotations}
Interestingly, the double Wick rotations discussed in \cite{Colleaux:2025uiw} do not affect the quantization, since the resulting metrics have the same reduced Lagrangians as the original ones. Consequently, the wave functions and CSs remain unchanged, so we only review the double Wick rotations of the symmetry-invariant metrics.

\subsubsection{B-metrics --- $[4,3,\{8,11\}]$}

\begin{itemize}
    \item In the case of s. Schwarzschild, performing the double Wick rotations $t=i\mathfrak{q}$ and $\theta = i \tilde{\theta}$, where $\rho = \sin \tilde{\theta}$, the symmetry-invariant metric is 
    \begin{equation}
        \check{g} =  q_1d\mathfrak{q}^2+ 2\tilde{q}_2d\mathfrak{q}dr+ q_3dr^2 +q_4\left(-\tfrac{d\rho^2}{\rho^2-1} + \rho^2 d\varphi^2\right),
    \end{equation}
after complex redefinition of the function $\tilde{q}_2=iq_2$. This metric has the symmetries of the BI-metric ($[4,3,8]$). Calculating $\varepsilon R$ and performing the symmetry reduction,
the reduced Lagrangian is the same as \eqref{spherical schwarzschild} with $k=1$. Then, it follows that the wave function and the CSs are the same as those of the s. Schwarzschild.
\item For h. Schwarzschild, after performing the transformations $\rho = \sqrt{\tanh ^2 h+p^2-1}\cosh h$ and\\ $\varphi= \cot ^{-1}\left({(p \sinh h)/\sqrt{p^2-1} }\right)$, and the double Wick rotations $t=i\mathfrak{q}$ and $h=i\psi$, we end up with the symmetry-invariant metric with the symmetries of BII-metric (also $[4,3,8]$):
\begin{equation}
    \check{g} =  q_1d\mathfrak{q}^2 + 2\tilde{q}_2d\mathfrak{q}dr + q_3 dr^2+ q_4\left( - p^2 d\psi^2 +\tfrac{d p^2}{p^2-1}\right),
\end{equation}
where $\tilde{q}_2=iq_2$.
The reduced Lagrangian from $\varepsilon R$ is given by \eqref{spherical schwarzschild} with $k=-1$. Thus, the wave function and the CSs are the same as those of the h. Schwarzschild metric.
\item For the p. Schwarzschild case, applying the double Wick rotations $t=i\mathfrak{q}$ and $\varphi=i\tanh ^{-1}\left(y/x\right)$, with $\rho^2=x^2-y^2$ yields the symmetry-invariant metric for the symmetries of the BIII-metric ($[4,3,11]$):
\begin{equation}
    \check{g} = q_1d\mathfrak{q}^2+ 2\tilde{q}_2 d\mathfrak{q}dr+ q_3dr^2+q_4\left(dx^2-dy^2\right),
\end{equation}
where $\tilde{q}_2 = i q_2$.
Calculating $\varepsilon R$ and following the previous prescription, we get the same reduced Lagrangian, thus, the same CSs and wave function, as p. Schwarzschild. 
\end{itemize}
\subsubsection{NHEK and swirling --- $[4,3,\{9,10\}]$}
\begin{itemize}
    \item For s. Taub--NUT, using the double Wick rotations $t=i\mathfrak{q}- 2n \varphi$ and taking the range of $\rho$ to be $(1,\infty)$,
    we get the symmetry-invariant metric with symmetries of NHEK ($[4,3,9]$):
    \begin{equation}
        \check{g} = q_1\left(d\mathfrak{q} - 2n\sqrt{\rho^2-1}d\varphi \right)^2+ 2 \tilde{q}_2 dr\left(d\mathfrak{q} -2n\sqrt{\rho^2-1}d\varphi\right) + q_3dr^2 + q_4\left(-\tfrac{d\rho^2}{\rho^2-1}+\rho^2d\varphi^2\right),
    \end{equation}
    where $\tilde{q}_2 = i q_2$.
    The reduced Lagrangian from the symmetry reduction using this invariant metric is the same as \eqref{general reduced lagrangian} with $k=1$. Therefore, the wave function and the (lack of) CSs are the same as for the original s. Taub--NUT.

    \item For h. Taub--NUT, the symmetry-invariant metric with symmetries of NHEK can be derived from the transformations $\rho = \sqrt{\tanh ^2 h+p^2-1}\cosh h, \, \varphi= \cot ^{-1}\left({(p \sinh h)/\sqrt{p^2-1} }\right)$ and \\$t=s + 2n \cot ^{-1}\left((p \sinh h)/\sqrt{p^2-1}\right)-2n \tan ^{-1}\left(\sqrt{p^2-1} \coth h\right)$, and the double Wick rotations $s=i\mathfrak{q}$ and $h=i\psi$, we end up with the symmetry-invariant metric 
    \begin{equation}
        \check{g} =  q_1\left(d\mathfrak{q} -2n \sqrt{p^2-1}d\psi \right)^2 + 2\tilde{q}_2 dr\left(d\mathfrak{q} -2n \sqrt{p^2-1}d\psi \right)+q_3dr^2 + q_4\left( - p^2 d\psi^2 +\tfrac{d p^2}{p^2-1}\right).
    \end{equation}
    where $\tilde{q}_2 = i q_2$.
    The reduced Lagrangian derived from the reduction of $\varepsilon R$ is the same as \eqref{general reduced lagrangian} with $k=-1$. Thus, the quantization and CSs are the same as the original h. Taub--NUT.

    \item For p. Taub--NUT, using the double Wick rotations $t=i(\mathfrak{q} +n xy)$, $\varphi=i\tanh ^{-1}\left(y/x\right)$ and the transformation $\rho=\sqrt{x^2-y^2}$, we get the symmetry-invariant metric with the symmetries of swirling universe ($[4,3,10]$):
    \begin{equation}
       \check{g} = q_1\left(d\mathfrak{q} +2nxdy \right)^2+ 2\tilde{q}_2dr\left(d\mathfrak{q} +2nxdy \right)+q_3dr^2+ q_4\left(dx^2-dy^2\right),
    \end{equation}
    where $\tilde{q}_2=i q_2$.
    The reduced Lagrangian from $\varepsilon R$ is the same as \eqref{general reduced lagrangian} with $k=0$ up to an overall constant of $1/2n$ that can be ignored. Thus, as in the previous cases, the quantization and the CSs are the same as p. Taub--NUT. 
\end{itemize}

\section{Symmetries of FLRW-type spacetimes --- $[6,3,-]$}\label{Quantization of FLRW spaces ($[6,3,-]$)}

Here, we consider the class of symmetry-invariant metrics possessing the symmetries of FLRW spacetimes ($[6,3,\{1,2,3\}]$), as well as the related symmetries $[6,3,\{4,5,6\}]$ where the orbit is $\mathrm{AdS}_3$, $\mathrm{M}_3$, or $\mathrm{dS}_3$, so we refer to them as the $\mathrm{AdS}_3/\mathrm{M}_3/\mathrm{dS}_3$-homogeneous spacetimes ($[6,3,\{4,5,6\}]$). 

\subsection{Closed, open, and flat FLRW --- $[6,3,\{1,3,2\}]$}\label{Quantization of $[6,3,123]$}
\subsubsection{Symmetry-invariant metric and classical solutions}
The Killing vectors generating the isometries of FLRW spacetimes depend on the spatial curvature. For closed FLRW spacetimes, they are
\begin{equation}
    \begin{aligned}
        X_1 &= \cos\varphi\partial_{\varv}-\cot \varv\sin\varphi\partial_{\varphi}, \quad X_2 = -\sin\varphi\partial_{\varv} - \cos\varphi\cot \varv\partial_{\varphi}, \quad X_3=\partial_{\varphi},\\
        X_4 &= \cos\varphi\sin \varv\partial_{\chi} + \cos \varv\cos\varphi\cot \chi\partial_{\varv} - \cot \chi\csc \varv\sin\varphi\partial_{\varphi},\\
        X_5&= \sin \varv\sin\varphi\partial_{\chi} + \cos \varv\cot \chi\sin\varphi\partial_{\varv} + \cos\varphi\cot \chi\csc \varv\partial_{\varphi},\\
        X_6&= \cos \varv\partial_{\chi} - \cot \chi\sin \varv\partial_{\varv}.
    \end{aligned}
\end{equation}
For open FLRW spacetimes, the Killing vectors read
\begin{equation}
    \begin{aligned}
        X_1 &= \cos\varphi\partial_{\varv}-\cot \varv\sin\varphi\partial_{\varphi}, \quad X_2 = -\sin\varphi\partial_{\varv} - \cos\varphi\cot \varv\partial_{\varphi}, \quad X_3=\partial_{\varphi},\\
        X_4 &= \cos\varphi\sin \varv\partial_{\chi} + \cos \varv\cos\varphi\coth \chi\partial_{\varv} - \coth \chi\csc \varv\sin\varphi\partial_{\varphi},\\
        X_5&= \sin \varv\sin\varphi\partial_{\chi} + \cos \varv\coth \chi\sin\varphi\partial_{\varv} + \cos\varphi\coth \chi\csc \varv\partial_{\varphi},\\
        X_6&= \cos \varv\partial_{\chi} - \coth \chi\sin \varv\partial_{\varv}.
    \end{aligned}
\end{equation}
Finally, for flat FLRW spacetimes, the Killing vectors take the simpler form
\begin{equation}
    \begin{aligned}
        X_1 &=\partial_{\varv},\quad X_2 = \partial_{\varphi}, \quad X_3 = -\varphi\partial_{\varv}+\varv \partial_{\varphi},\quad X_4 = \varphi \partial_{\chi}-\chi \partial_{\varphi}, \quad X_5 = \varv \partial_{\chi} - \chi \partial_{\varv},\quad X_6 = -\partial_{\chi}.
    \end{aligned}
\end{equation}
The symmetry-invariant metrics are
\begin{equation}\label{FLRW metric}
  \check{g} = -q_1(t)^2dt^2 + q_2(t)^2 \begin{cases}
        d\chi^2+\sin^2 \chi(d\varv^2+ \sin^2 \varv d\varphi^2),\\
        d\chi^2+\sinh^2 \chi(d\varv^2+ \sin^2 \varv d\varphi^2),\\
        d\chi^2+d\varv^2+d\varphi^2,\\
    \end{cases}  
\end{equation}
for closed, open, and flat FLRW spacetimes, respectively. Note that the most general symmetry-invariant metric uses $q_1(t)$ and $q_2(t)$ without squaring, due to the linearity of $\mathcal{L}_{X_i}\check{g}=0$. However, in this form, the supermetric and superpotential become constant, causing the quantization method used in this paper to break down. Hence, we retain the squares in the symmetry-invariant metrics. The only symmetry-invariant metrics that admit a classical vacuum solution for GR (up to gauge transformations) correspond to the p. FLRW with ${q_1(t)=1}$ and ${q_2(t)=\textrm{const.}}$ or the open FLRW with ${q_1(t)=1}$ and ${q_2(t)=t}$ both of which describe (parts of) the Minkowski spacetime. In the following, we construct the reduced Lagrangian for the minisuperspaces corresponding to the symmetry-invariant metrics \eqref{FLRW metric}, determine their CSs, and subsequently proceed to their quantization.

\subsubsection{Reduced Lagrangian and generalized momenta}

To derive the reduced Lagrangian, we substitute the symmetry-invariant metric into $\varepsilon R$. This yields
\begin{equation}
\varepsilon R =
\begin{cases}
  \frac{6 q_2 \left(-q_2 \dot{q}_2 \dot{q}_1+q_1 \left(q_2 \ddot{q}_2+\dot{q}_2^2\right)+ q_1^3\right)}{q_1^2}\sin \varv \, \sin^2 \chi,\\
   \frac{ 6q_2 \left(-q_2 \dot{q}_2 \dot{q}_1+q_1 \left(q_2 \ddot{q}_2+\dot{q}_2^2\right)- q_1^3\right)}{q_1^2}\sin \varv \, \sinh^2 \chi,\\
 \frac{6 q_2 \left(-q_2 \dot{q}_2 \dot{q}_1+q_1 \left(q_2 \ddot{q}_2+\dot{q}_2^2\right)\right)}{q_1^2},
\end{cases}
\end{equation}
where $\dot{q}_1 = \tfrac{dq_1}{dt}$ and $\dot{q}_2=\tfrac{dq_2}{dt}$. 
The reduced Lagrangians for the three cases can be written in the compact form
\begin{equation}
    \check{L} = 6q_2\left(kq_1-\tfrac{\dot{q}_2^2}{q_1}\right),
\end{equation}
where $k=1,-1,0$ represent closed, open, and flat FLRW, respectively.
Writing the reduced Lagrangian in the form $\check{L} = \tfrac{1}{2q_1}G_{\alpha\beta}(q_2)\dot{q}^{\alpha}\dot{q}^{\beta}-q_1V(q_2)$, the supermetric $G_{\alpha\beta}$ in this case is a $1\times 1$ matrix: $G_{\alpha\beta}(q_2) = -12q_2$, with $\alpha,\beta =2$ and $V(q_2)=-6kq_2$. The canonical momentum conjugate to $q_2$ is derived from $p_2 = \tfrac{\partial \check{L}}{\partial \dot{q}_2}$, resulting in
\begin{equation}
    p_2 = -12\frac{q_2\dot{q}_2}{q_1},
\end{equation}
thus, $\dot{q}_2 = - \tfrac{q_1 p_2}{12q_2}$. Using these formulae, the Hamiltonian $H = \dot{q}_2p_2-\check{L}$ is given by
\begin{equation}
    H = q_1 \left(- \tfrac{p^2_2}{24q_2}-6kq_2\right).
\end{equation}

\subsubsection{Conditional symmetries}

The generators of the CS algebra are easily derivable from
$\pounds_{\xi}G^{\alpha\beta}=\Phi G^{\alpha\beta}$ and $\pounds_{\xi}V=\Phi V$, where $\Phi=\Phi(q_2)$, $\alpha,\beta = q_2$ and $\xi^{\alpha}(q_2)$ is the CS generator and is composed of a component which we will call $\xi$. These two equations converge into $ \xi k = q_2 \Phi k$. If $k=\pm 1$, then $\xi = q_2\Phi$. If $k=0$, the equation $\pounds_{\xi}V=\Phi V$ is satisfied trivially and from $\pounds_{\xi}G^{\alpha\beta}=\Phi G^{\alpha\beta}$, we also end up with $\xi = q_2 \Phi$. The COMs are given by $Q = \xi p_2 = q_2\Phi p_2$. From the requirement $\{Q,H\} \propto H$, we get $\Phi= \tfrac{1}{q_2}$, i.e., $\xi=1$ and $Q=p_2$. Thus, the algebra of CSs is trivial.

\subsubsection{Quantization}

Moving on to the quantization procedure, we promote the coordinates and momenta to operators similarly to the previous sections. The WDW equation is
\begin{equation}
    \tfrac{1}{24q_2}\tfrac{d^2\Psi}{dq_2^2} - \tfrac{1}{48q_2^2}\tfrac{d\Psi}{dq_2}-6kq_2\Psi = 0.
\end{equation}
For $k=\pm 1$, the general solution to this equation is given by
\begin{equation}\label{WF of [6,3,{1,3}]}
    \Psi(q_2) = 3^{3/8}(-k)^{3/16}q_2^{3/4}\Big[A J_{-\tfrac{3}{8}}\left(6i\sqrt{k}q_2^2\right)\Gamma\left(\tfrac{5}{8}\right)+BJ_{\tfrac{3}{8}}\left(6i\sqrt{k}q_2^2\right)\Gamma(\tfrac{11}{8})\Big],
\end{equation}
where $A$ and $B$ are arbitrary (normalization) complex constants,
while for $k=0$, the general solution is
\begin{equation}\label{WF for [6,3,2]}
    \Psi(q_2) = Aq_2^{3/2}+B.
\end{equation}

\subsection{$\mathrm{AdS_3}$-homogeneous --- $[6,3,4]$}

In this case, the symmetry-invariant metric given by
\begin{equation}
    \check{g} = q_1(t)^2 dt^2+q_2(t)^2\left(-\cosh^2 y_2dy_1^2+dy_2^2+\sinh^2y_2dy_3^2\right).
\end{equation}
The corresponding reduced Lagrangian is derived as usual with the end result being
\begin{equation}
    \check{L} = \tfrac{6q_2\dot{q}_2^2}{q_1}-6q_2q_1,
\end{equation}
which is the same as the reduced Lagrangian of $[6,3,1]$ up to an overall sign. This implies that the Hamiltonian coincides with that of $[6,3,1]$, namely $H = -\tfrac{p_2^2}{24q_2} - 6q_2$. Upon quantization, the WDW equation and the CSs remain unchanged. Thus, the wave function for $[6,3,4]$ is given by \eqref{WF of [6,3,{1,3}]} with $k=1$.

\subsection{$\mathrm{dS_3}$-homogeneous --- $[6,3,6]$}

In this case, the symmetry-invariant metric is
\begin{equation}
    \check{g} = q_1(t)^2dt^2+q_2(t)^2\left(-dy_1^2+\cosh^2 y_1(dy_2^2+\sin^2 y_2dy_3^2)\right).
\end{equation} 
The corresponding reduced Lagrangian reads
\begin{equation}
    \check{L} = -6q_2\dot{q}_2^2-6q_2,
\end{equation}
which matches that of $[6,3,3]$ up to an overall sign. Consequently, the Hamiltonian is $H = - \tfrac{p_2^2}{24q_2}+6q_2$, and the quantization yields the same WDW equation and CSs as for $[6,3,3]$. The corresponding wave function is therefore given by \eqref{WF of [6,3,{1,3}]} with $k=-1$.

\subsection{$\mathrm{M_3}$-homogeneous --- $[6,3,5]$}

Here, the symmetry-invariant metric is given by
\begin{equation}
    \check{g} = q_1(t)^2dt^2+q_2(t)^2\left(-dy_1^2+dy_2^2+dy_3^2\right).
\end{equation}

The reduced Lagrangian is
\begin{equation}
    \check{L} = -6q_2\dot{q}_2^2,
\end{equation}
identical to that of $[6,3,2]$. Accordingly, the Hamiltonian coincides with the $[6,3,2]$ case, $H=-\tfrac{p_2^2}{24q_2}$, and the WDW equation and CSs are the same. The resulting wave function is therefore given by \eqref{WF for [6,3,2]}.

\subsection{Models with cosmological constant and scalar field}\label{Adding a scalar field and a cosmological constant}
As mentioned in the beginning of the section, only flat or open FLRW admit a vacuum solution in GR and this is just the Minkowski spacetime which belongs to $[10,4,-]$ instead (i.e., maximally symmetric spacetime). In order to quantize models with solutions that belong strictly to $[6,3,-]$, we add a cosmological constant and a massless scalar field. The cosmological-constant and matter terms to be added to the Lagrangian can be written as
\begin{equation}\label{eq:matter}
\check{L}_{\text{matter}}= \epsilon\left(\Lambda + \partial_{\mu}\phi\partial^{\mu}\phi\right) ,
\end{equation}
where $\Lambda$ is the cosmological constant and $\phi=\phi(t)$ is a massless scalar field.

\subsubsection{Reduced Lagrangian and generalized momenta}

Substituting the symmetry-invariant metrics \eqref{FLRW metric} in \eqref{eq:matter} and adding these terms to the gravitational reduced Lagrangian, we get the total reduced Lagrangian
\begin{equation}
    \check{L} = \frac{1}{q_1}\left(-6q_2\dot{q}_2^2 + q_2^3 \dot{\phi}^2\right)+q_1 \left(6kq_2+\Lambda q_2^3\right).
\end{equation}
This expression can be recast to the form $\check{L} = \tfrac{1}{2q_1}G_{\alpha\beta}(q)\dot{q}^{\alpha}\dot{q}^{\beta}-q_1V(q)$, where $q=(q_2,\phi)$ and the supermetric and superpotential read
\begin{equation}
    \begin{aligned}
        G_{\alpha\beta} = \begin{pmatrix}
            -12q_2&0\\0&2q_2^3
        \end{pmatrix}, \quad V=-6kq_2-\Lambda q_2^3.
    \end{aligned}
\end{equation}
The canonical momenta are given by
\begin{equation}
    p_2 = \tfrac{\partial \check{L}}{\partial \dot{q}_2} = -12\frac{q_2\dot{q}_2}{q_1}, \quad p_{\phi}= \tfrac{\partial \check{L}}{\partial \dot{\phi}}=2\frac{q_2^3\dot{\phi}}{q_1}.
\end{equation}
Using
\begin{equation}
    \dot{q}_2=-\tfrac{q_1 p_2}{12q_2}, \quad \dot{\phi}=\tfrac{q_1 p_{\phi}}{2q_2^3},
\end{equation}
the Hamiltonian becomes
\begin{equation}
    H = p_2\dot{q}_2+p_{\phi}\dot{\phi}-\check{L} = q_1 \left(-\tfrac{p_2^2}{24q_2}+\tfrac{p_{\phi}^2}{4q_2^3}-6kq_2-\Lambda q_2^3\right).
\end{equation}

\subsubsection{Conditional symmetries}

The generators of the CS algebra $\xi^{\alpha}=(\xi^1(q),\xi^4(q))$ are calculated from
\begin{equation}
    \pounds_{\xi}G^{\alpha\beta}= \Phi  G^{\alpha\beta}, \quad \pounds_{\xi}V= \Phi V,
\end{equation}
where $\Phi= \Phi(q)$.
For $k=\pm 1$, the equations reduce to
\begin{equation}
    \begin{aligned}
&q_2^2\partial_{q_2}\xi^4=6\partial_{\phi}\xi^1, \quad \xi^1+2q_2\partial_{q_2}\xi^1 = -\tfrac{\xi^1 (6k+3\Lambda q_2^2)}{6k+\Lambda q_2^2}, \quad 3\xi^1+2q_2\partial_{\phi}\xi^4=- \tfrac{\xi^1(6k+3\Lambda q_2^2)}{6k+\Lambda q_2^2}. 
    \end{aligned}
\end{equation}
The solution is $\xi^1=0$ and $\xi^4=A$, where $A$ is a real constant. Thus, the generator for the $1d$ CS algebra is $\xi^{\alpha} = (0,1)$, which yields the constant of motion $Q=\xi^{\alpha}p_{\alpha}=p_{\phi}$.
 
 For $k=0$, the generators of the CS algebra are derived by solving
 \begin{equation}
     \begin{aligned}
         q_2^2 \partial_{q_2}\xi^4=6\partial_{\phi}\xi^1,\quad q_2\partial_{q_2}\xi^1=-2\xi^1, \quad q_2^2\partial_{\phi}\xi^4=-3\xi^1.
     \end{aligned}
 \end{equation}
The general solution is 
\begin{equation}
    \xi^1=\tfrac{A}{q_2^2}e^{\sqrt{3/2}\phi}+ \tfrac{B}{q_2^2}e^{-\sqrt{3/2}\phi} ,\quad \xi^4=-\tfrac{\sqrt{6}A}{q_2^3}e^{\sqrt{3/2}\phi}+ \tfrac{\sqrt{6}B}{q_2^3}e^{-\sqrt{3/2}\phi}+C,
\end{equation}
where $A$, $B$, and $C$ are real constants. Thus, the generators of the corresponding $3d$ CS algebra are given by
\begin{equation}
    \xi_1=(0,1), \quad \xi_2=\left(\tfrac{e^{-\sqrt{3/2}\phi}}{q_2^2},\sqrt{6}\tfrac{e^{-\sqrt{3/2}\phi}}{q_2^3}\right), \quad \xi_3=\left(\tfrac{e^{\sqrt{3/2}\phi}}{q_2^2},-\sqrt{6}\tfrac{e^{\sqrt{3/2}\phi}}{q_2^3}\right).
\end{equation}
The COMs derived from these generators read $Q_{1}=p_{\phi}$, $Q_{2} = \tfrac{e^{-\sqrt{3/2}\phi}}{q_2^2}p_2 + \sqrt{6}\tfrac{e^{-\sqrt{3/2}\phi}}{q_2^3}p_{\phi}$, and $Q_3 = \tfrac{e^{\sqrt{3/2}\phi}}{q_2^2} p_2-\sqrt{6}\tfrac{e^{\sqrt{3/2}\phi}}{q_2^3}p_{\phi}$, with Poisson brackets $\{Q_1,Q_2\}=Q_2$, $\{Q_3,Q_1\}=Q_3$, and $\{Q_3,Q_2\}=0$. Again, the resulting algebra is of Bianchi type VI with $h=-1$.

\subsubsection{Quantization}

Apart from the operators $\hat{q}_2$ and $\hat{p}_2$ we also introduce $\phi\to \hat{\phi}$ such that $\hat{\phi}\ket{\Psi}=\phi\ket{\Psi}$ and $\hat{p}_{\phi}\ket{\Psi} = -i\partial_{\phi}\ket{\Psi}$. The WDW equation is
\begin{equation}
    \left(-\tfrac{1}{2\mu}\partial_{\alpha}\left(\mu G^{\alpha\beta}\partial_{\beta}\right)+V\right)\Psi(q)=0.
\end{equation}
Upon inserting the supermetric and superpotential, it takes the form
\begin{equation}\label{WDW eq. for FLRW with scalar field}
    \tfrac{q_2^2}{6}\partial_{q_2}^2\Psi+\tfrac{q_2}{6}\partial_{q_2}\Psi+\partial_{\phi}^2\Psi+(24kq_2^2+4\Lambda q_2^4)\Psi=0.
\end{equation}
From this we distinguish between two cases: 
\begin{enumerate}
    \item $k=\pm 1$

The general solution of \eqref{WDW eq. for FLRW with scalar field} is
\begin{equation}\label{WF for [6,3,1] with matter}
\begin{aligned}
\Psi(q)
&= q_2^{-1} 
e^{-\sqrt{6\Lambda}q_2^{2}}
\big( 2\sqrt{6\Lambda}q_2^{2} \big)^{\frac{1}{2}\sqrt{1+6\lambda} + \tfrac{1}{2}}
\Big[ A\cos(\sqrt{\lambda}\phi) + B\sin(\sqrt{\lambda}\phi) \Big] \\
&\times
\Bigg\{
C_{1}
{}_1F_1\left(
\tfrac{1}{2}\sqrt{1+6\lambda}
+ \tfrac{18k}{\sqrt{6\Lambda}}
+ \tfrac{1}{2};
\sqrt{1+6\lambda} + 1;
2\sqrt{6\Lambda}q_2^{2}
\right) \\
&
+ C_{2}
\big( 2\sqrt{6\Lambda}q_2^{2} \big)^{-\sqrt{1+6\lambda}}
{}_1F_1\left(
-\tfrac{1}{2}\sqrt{1+6\lambda}
+ \tfrac{18k}{\sqrt{6\Lambda}}
+ \tfrac{1}{2};
1 - \sqrt{1+6\lambda};
2\sqrt{6\Lambda}q_2^{2}
\right)
\Bigg\},
\end{aligned}
\end{equation}
where ${}_1F_1(a;b;z)$ is the confluent hypergeometric function of the first kind, $A,B,C_1,C_2$ and $\lambda$ are arbitrary complex constants.

We can simplify the equation using the CSs, which is promoted to an operator $Q\to \hat{Q}$ such that $\hat{Q}\ket{\Psi}=\kappa \ket{\Psi}$ with $\kappa$ being a constant. Since $Q=p_{\phi}$, we have $\hat{Q}\ket{\Psi} = \hat{p}\ket{\Psi}=-i\partial_{\phi}\ket{\Psi}$. This imposes an additional eigenvalue equation $ -i\partial_{\phi}\Psi=\kappa\Psi$  whose solution is $\Psi(q)= \tilde{\Psi}(q_2)e^{i\kappa\phi}$. Substituting into the WDW equation, we get
\begin{equation}\label{Q_1 WDW eq} 
   \tfrac{q_2^2}{6}\partial_{q_2}^2\tilde{\Psi}+\tfrac{q_2}{6}\partial_a\tilde{\Psi}+\left(24kq_2^2+4\Lambda q_2^4-\kappa^2\right)\tilde{\Psi}=0. 
\end{equation}
The solution is
\begin{equation}\label{Q_1 solution}
\begin{aligned}
\tilde{\Psi}(q_2) 
&= C_{1}(6\Lambda)^{\frac{1+\sqrt{1-\frac{3}{2}\kappa^{2}}}{4}} 
q_2^{1+\sqrt{1-\frac{3}{2}\kappa^{2}}}
e^{-\tfrac{1}{2}\sqrt{6\Lambda}q_2^{2}}
{}_1F_1\left(
\tfrac{1}{2}\left(1+\sqrt{1-\tfrac{3}{2}\kappa^{2}}\right)
+\tfrac{9k}{\sqrt{6\Lambda}};
1+\sqrt{1-\tfrac{3}{2}\kappa^{2}};
\sqrt{6\Lambda}q_2^{2}
\right)\\& + 
C_{2}(6\Lambda)^{\frac{1-\sqrt{1-\frac{3}{2}\kappa^{2}}}{4}} 
q_2^{1-\sqrt{1-\frac{3}{2}\kappa^{2}}}
e^{-\tfrac{1}{2}\sqrt{6\Lambda}q_2^{2}}
{}_1F_1\left(
\tfrac{1}{2}\left(1-\sqrt{1-\tfrac{3}{2}\kappa^{2}}\right)
+\tfrac{9k}{\sqrt{6\Lambda}};
1-\sqrt{1-\tfrac{3}{2}\kappa^{2}};
\sqrt{6\Lambda}q_2^{2}
\right),
\end{aligned}
\end{equation}
where $C_1$ and $C_2$ are arbitrary complex constants.
Then, the wave function is given by
\begin{equation}
    \Psi(q)=\tilde{\Psi}(q_2)e^{i\kappa \phi}.
\end{equation}

\item $k=0$

    The general solution of \eqref{WDW eq. for FLRW with scalar field} is given by
    \begin{equation}\label{WF for [6,3,2] with matter}
\begin{aligned}
\Psi(q) &= 
C_1 q_2^{1+\sqrt{1-\frac{3}{2}\lambda^2}}
e^{-\frac{1}{2}\sqrt{6\Lambda}q_2^2}
{}_1F_1\Bigg(
\tfrac{1+\sqrt{1-\frac{3}{2}\lambda^2}}{2};1+\sqrt{1-\tfrac{3}{2}\lambda^2}; \sqrt{6\Lambda}q_2^2
\Bigg) e^{i \lambda \phi} \\
&+ C_2 q_2^{1-\sqrt{1-\frac{3}{2}\lambda^2}} 
e^{-\frac{1}{2}\sqrt{6\Lambda}q_2^2}
{}_1F_1\Bigg(
\tfrac{1-\sqrt{1-\frac{3}{2}\lambda^2}}{2}; 1-\sqrt{1-\tfrac{3}{2}\lambda^2};\sqrt{6\Lambda} q_2^2
\Bigg) e^{-i \lambda \phi},
\end{aligned}
\end{equation}
where $C_1$ and $C_2$ are arbitrary complex constants.
As in the previous case, we can simplify the Hamiltonian constraint using CSs. Upon quantization, the eigenvalue equations for the CSs are 
\begin{equation}
    -i\partial_{\phi}\Psi=\kappa_1\Psi, \quad -i\left(\tfrac{e^{-\sqrt{3/2}\phi}}{q_2^2}\partial_{q_2} + \sqrt{6}\tfrac{e^{-\sqrt{3/2}\phi}}{q_2^3}\partial_{\phi}\right)\Psi=\kappa_2 \Psi, \quad -i\left(\tfrac{e^{\sqrt{3/2}\phi}}{q_2^2} \partial_{q_2}-\sqrt{6}\tfrac{e^{\sqrt{3/2}\phi}}{q_2^3}\partial_{\phi}\right)\Psi=\kappa_3\Psi,
\end{equation}
where $\kappa_1$, $\kappa_2$, and $\kappa_3$ are constants. Imposing all equations leads to a vanishing wave function ($\Psi(q)=0$). Restricting ourselves to subalgebras of the CS algebra, we find that all subalgebras lead to trivial wave functions except $\langle Q_1 \rangle$ which results in \eqref{Q_1 solution} with $k=0$ and $\kappa \to \kappa_1$.
\end{enumerate}
As explained in the previous section, the CSs and the quantization of the $\mathrm{AdS_3}$/$\mathrm{M_3}$/$\mathrm{dS_3}$-homogeneous spacetimes $[6,3,\{4,5,6\}]$ with a cosmological constant and a massless scalar field yield the same results as those of $[6,3,\{1,2,3\}]$ explored in this section.

Above, we reviewed the quantization of FLRW cosmologies in vacuum and with an added cosmological constant and a minimally coupled massless scalar field. It is worth mentioning that these models were studied extensively with different matter content. The minisuperspace quantization of FLRW was performed with a minimally coupled scalar field \cite{Lin:2023uxq,Dimakis:2016mip}. Adding an aether to the scalar field was done in \cite{Dimakis:2020zbb}.
The quantization of FLRW with $n$ minimally coupled scalar fields was carried out in \cite{Kim:2012js}, and with a conformally coupled scalar field with a cosmological constant was discussed in \cite{Kocher:2018ilr}. Dust and radiation (as two decoupled fluids) were added to FLRW cosmology in \cite{Pinto-Neto:2005pwn}, and an electromagnetic field in \cite{Jalalzadeh:2022bgz}.
The most general scalar-tensor theory was considered in \cite{Borowiec:2021tyg}. Supersymmetry was discussed in \cite{Ramirez:2015vsj}.

\section{Symmetries of Bianchi class A --- $[3,3,\{2,3,8,9\}]$}\label{Sec.[3,3,-]}

Bianchi class A models refer to Bianchi types I, II, VIII, IX, which are all PSC-compatible. In contrast, the Bianchi class B models do not satisfy PSC2, and their minisuperspaces, therefore lead to incorrect reduced field equations \cite{Hawking:1968zw,Fels:2001rv}. In GR, only the most general Bianchi type I and II models admit exact vacuum solutions strictly with such symmetries. In contrast, for Bianchi type VIII and IX models, exact solutions exist only with enhanced symmetries, as we will see below.

\subsection{Bianchi type I --- $[3,3,2]$}
In this section, following \cite{Christodoulakis:2001um}, we review the quantization of symmetry-invariant metrics with Bianchi type I symmetries.

\subsubsection{Symmetry-invariant metric and classical solutions}
The Killing vectors defining the symmetries of Bianchi type I ($[3,3,2]$) are
\begin{equation}
    X_1=\partial_{x^1},\quad X_2=\partial_{x^2}, \quad  X_3=\partial_{x^3}.
\end{equation}
The general symmetry-invariant metric reads
\begin{equation}\label{eq:Bianchimetric}
    \check{g} = (-N(t)+N_{\alpha}(t)N^{\alpha}(t))dt^2 + 2 N_{\alpha}(t)\sigma^{\alpha}(x)dt + \gamma_{\alpha\beta}\sigma^{\alpha}(x)\sigma^{\beta},
\end{equation}
with $\sigma^{\alpha}(x)$ being the 1-forms satisfying $d\sigma^{\alpha}(x)=0$, i.e., $\sigma^{\alpha} = dx^{\alpha}$, 
and $\gamma_{\alpha\beta}$ being the Riemannian metric on the homogeneous $3d$ space. 
A known vacuum GR solution strictly within this class is the Kasner cosmology \cite{Kasner:1921zz},
\begin{equation}
    ds^2 = -dt^2 + t^{2p_1}(dx^1){}^2+t^{2p_2}(dx^2){}^2+t^{2p_3}(dx^3){}^2, \quad \sum_{\mathclap{i=1}}^3p_i=1.
\end{equation}

\subsubsection{Hamiltonian and generalized momenta}

Following \cite{GESneddon_1976}, the Hamiltonian can be written as
\begin{equation}
    H = \tfrac{N}{\sqrt{\gamma}}H_0,
\end{equation}
where $\gamma = \det(\gamma_{\alpha\beta})$. For Bianchi type I, we have
\begin{equation}
    H_0 = \tfrac{1}{2}L_{\alpha\beta\mu\nu}\pi^{\alpha\beta}\pi^{\mu\nu}, 
\end{equation}
where  $L_{\alpha\beta\mu\nu} = \gamma_{\alpha\mu}\gamma_{\beta\nu}+\gamma_{\alpha\nu}\gamma_{\beta\mu}-\gamma_{\alpha\beta}\gamma_{\mu\nu}$, and $\pi^{\mu\nu}$ are the canonical momenta dual to $\gamma_{\mu\nu}$ in the sense that they satisfy 
\begin{equation}\label{poisson bracket of gamma and pi}
    \{\gamma_{\alpha\beta},\pi^{\mu\nu}\}= \delta_{\alpha}^{\mu}\delta_{\beta}^{\nu}-\delta_{\beta}^{\mu}\delta_{\alpha}^{\nu}.
\end{equation}
\subsubsection{Conditional symmetries}
In this formalism, the generators of the CS algebra are given by \cite{Christodoulakis:2001um}
\begin{equation}
    E_{(I)}{}_{\beta\rho} = \lambda_{(I)\beta}^{\alpha}\gamma_{\alpha\rho},
\end{equation}
where $I = \{1,2,\cdots,9\}$ and 
\begin{equation}\label{generators for [3,3,1]}
    \begin{aligned}
&\lambda_1 =
\begin{pmatrix}
0 & 1 & 0 \\
0 & 0 & 0 \\
0 & 0 & 0
\end{pmatrix}, \quad
\lambda_2 =
\begin{pmatrix}
0 & 0 & 1 \\
0 & 0 & 0 \\
0 & 0 & 0
\end{pmatrix}, \quad
\lambda_3 =
\begin{pmatrix}
0 & 0 & 0 \\
0 & 0 & 1 \\
0 & 0 & 0
\end{pmatrix},\quad
\lambda_4 =
\begin{pmatrix}
0 & 0 & 0 \\
0 & 0 & 0 \\
0 & 1 & 0
\end{pmatrix}, \quad
\lambda_5 =
\begin{pmatrix}
0 & 0 & 0 \\
0 & 0 & 0 \\
1 & 0 & 0
\end{pmatrix},\\&
\lambda_6 =
\begin{pmatrix}
0 & 0 & 0 \\
1 & 0 & 0 \\
0 & 0 & 0
\end{pmatrix},\quad
\lambda_7 =
\begin{pmatrix}
1 & 0 & 0 \\
0 & -1 & 0 \\
0 & 0 & 0
\end{pmatrix}, \quad
\lambda_8 =
\begin{pmatrix}
0 & 0 & 0 \\
0 & -1 & 0 \\
0 & 0 & 1
\end{pmatrix}, \quad
\lambda_9 =
\begin{pmatrix}
1 & 0 & 0 \\
0 & 1 & 0 \\
0 & 0 & 1
\end{pmatrix}.
    \end{aligned}
\end{equation}
The COMs are then given by
\begin{equation}
    Q_{(I)} = E_{(I)}{}_{\beta\rho}\pi^{\beta\rho},
\end{equation}
which can be shown to satisfy $\{Q_{(I)},H_0\}=0$. Notice that the matrices $\lambda_{(I)}$ form an algebra
\begin{equation}
    [\lambda_{(I)},\lambda_{(J)}]=C_{IJ}^M\lambda_{(M)},
\end{equation}
implying that $\{Q_{(I)},Q_{(J)}\}=-\tfrac{1}{2}C_{IJ}^MQ_{(M)}$,
with the non-vanishing structure constants $C_{IJ}^M$ given by
\begin{equation}\label{structure constants for [3,3,1]}
    \begin{aligned}
&C_{13}^2=C_{24}^1=C_{28}^2=C_{46}^5=C_{25}^7=C_{68}^6=C_{16}^7=C_{35}^6=C_{37}^3=C_{57}^5=1, \quad C_{67}^6=C_{38}^3=2, \quad C_{48}^4=C_{17}^1=-2\\&
C_{15}^4=C_{34}^8=C_{47}^4=C_{25}^8=C_{18}^1=C_{26}^3=C_{27}^2=C_{58}^5=-1.
    \end{aligned}
\end{equation}
\subsubsection{Quantization}
To quantize this system, we promote $\gamma_{\alpha\beta}$ and $\pi^{\alpha\beta}$ to operators as follows: $\gamma_{\alpha\beta} \to \hat{\gamma}_{\alpha\beta}$ such that 
$\hat{\gamma}_{\alpha\beta}\ket{\Psi}=\gamma_{\alpha\beta}\ket{\Psi}$ and $\pi^{\alpha\beta} \to 
\hat{\pi}^{\alpha\beta}$ such that $\hat{\pi}^{\alpha\beta}\ket{\Psi} =-i \tfrac{\partial }{\partial 
\gamma_{\alpha\beta}}\ket{\Psi}$. The commutators between $\hat{\gamma}_{\alpha\beta}$ and $\hat{\pi}^{\alpha\beta}$ follow from \eqref{poisson bracket of gamma and pi} by replacing the Poisson bracket with a commutator:
\begin{equation} [\hat{\gamma}_{\alpha\beta},\hat{\pi}^{\mu\nu}] = \tfrac{i}{2}(\delta_{\alpha}^{\mu}\delta_{\beta}^{\nu}-\delta_{\beta}^{\mu}\delta_{\alpha}^{\nu}).
\end{equation}
The eigenvalue equations for the COMs are given by
\begin{equation}\label{eigenvalue eq. for bianchi I}
    \hat{Q}_{(I)}\hat{\Psi} = -i \lambda_{(I)\alpha}^{\tau}\gamma_{\tau\beta}\tfrac{\partial \ket{\Psi}}{\partial \gamma_{\alpha\beta}}=\kappa_{(I)}\ket{\Psi},
\end{equation}
where $\kappa_{(I)}$ are constants. The commutators between $\hat{Q}_{I}$s are given by replacing the Poisson bracket in the classical expression with a commutator, i.e., $[\hat{Q}_{I},\hat{Q}_{(J)}] = -\tfrac{1}{2}C_{IJ}^M\hat{Q}_{(M)}$, i.e., the COM algebra is $9-$dimensional and non-Abelian. Applying these to $\ket{\Psi}$, we get the selection rule $C_{IJ}^M\kappa_{(M)}=0$, and given the values of the structure constants, we deduce that $\kappa_i=0$ where $i\in \{1,2,\cdots,8\}$, but $\kappa_9$ remains arbitrary. Solving the system of equations \eqref{eigenvalue eq. for bianchi I} in light of the values of $\kappa_{(I)}$, we get
\begin{equation}\label{sol. for Bianchi I}
    \Psi(\gamma) = c \gamma^{i\kappa_9/3},
\end{equation}
where $c$ is an arbitrary constant, that is, the wave function $\Psi$ is only a function of the determinant of the spatial metric $\gamma_{\alpha\beta}$. To fix $\kappa_9$, we need the Hamiltonian constraint/WDW equation $\hat{H}_0\ket{\Psi} = 0$. Promoting $H_0$ into an operator naively will result in the usual ordering problem where the operator is not Hermitian; instead, we rearrange the terms in $H_0$ in such a way that its corresponding operator is Hermitian. The correct formula for the Hamiltonian operator is found from the identity (for the full derivation, see appendix A in \cite{Christodoulakis:2001um})
\begin{equation}\label{H_0 operator}
L_{\alpha\beta\mu\nu}\pi^{\alpha\beta}\pi^{\mu\nu}=\left(L_{\alpha\beta\mu\nu}\tfrac{\partial^2}{\partial \gamma_{\alpha\beta}\partial\gamma_{\mu\nu}}+3\gamma_{\kappa\lambda}\tfrac{\partial}{\partial \gamma_{\kappa\lambda}}+ \tfrac{3}{2}\right).
\end{equation}
The Hamiltonian constraint then reads
\begin{equation}
    -\tfrac{1}{2}\left(L_{\alpha\beta\mu\nu}\tfrac{\partial^2\Psi}{\partial \gamma_{\alpha\beta}\partial\gamma_{\mu\nu}}+3\gamma_{\kappa\lambda}\tfrac{\partial \Psi}{\partial \gamma_{\kappa\lambda}}+ \tfrac{3}{2}\Psi\right)=0.
\end{equation}
This equation, by itself, cannot be solved analytically, however, imposing the CSs, i.e., substituting \eqref{sol. for Bianchi I}, we get $\kappa_9 = \pm \tfrac{3i}{\sqrt{2}}$. Thus, the wave function for the quantized vacuum Bianchi type I is given by
\begin{equation}\label{WF for [3,3,1]}
    \Psi(\gamma) =  c_1 \gamma^{1/\sqrt{2}}+c_2\gamma^{-1/\sqrt{2}},
\end{equation}
where $c_1$ and $c_2$ are arbitrary constants.

In addition to the vacuum case, versions including matter have also been quantized in the literature. For instance, Bianchi type I cosmology with $n$ scalar fields and an exponential potential was quantized in \cite{Socorro:2019wpu}, the inclusion of a radiation field was studied in \cite{SHEN1997389}, and the model with toroidal spatial sections and a cosmological constant was analyzed in \cite{Hervik:2000ed}.

\subsection{Bianchi type II --- $[3,3,3]$}
Now, we follow \cite{Christodoulakis:2001de,Christodoulakis:2000xe,CHRISTODOULAKIS199755} reviewing the quantization of the symmetry-invariant metrics possessing the symmetries of Bianchi type II ($[3,3,3]$).

\subsubsection{Symmetry-invariant metric and classical solutions}
The Killing vectors defining the symmetries of Bianchi type II model are
\begin{equation}
    \begin{aligned}
        X_1 = \partial_{x^1}, \quad X_2 = \partial_{x^4}, \quad X_3 = x^4\partial_{x^1}+\partial_{x^3}.
    \end{aligned}
\end{equation}
The symmetry-invariant metric is given by \eqref{eq:Bianchimetric}, where now $\sigma^1(x)=dx^2-x^1dx^3$, $\sigma^2(x)=dx^3$, and $\sigma^3(x)=dx^1$ satisfying $d\sigma^{\alpha}(x)=\tfrac{1}{2}C^{\alpha}_{\beta\gamma} \sigma^{\beta}\wedge \sigma^{\gamma}$ with $C_{23}^1=-C_{32}^1=1$. 
The general Bianchi type II admits, for example, the vacuum GR solution from \cite{Taub:1950ez,Lorenz:1980fk}:
\begin{equation}
    ds^2 = -dt^2 + f_1(t)(\sigma^1)^2 + f_2(t)\left((\sigma^2)^2+(\sigma^3)^2\right),
\end{equation}
with $f_1(\tau)^2= \tfrac{2 a_1}{\cosh(2a_1 \tau + a_2)}$, $f_2(\tau)^2 = \tfrac{e^{2 a_1 \tau + a_3}}{2a_1}\cosh(2a_1 \tau + a_2)$, where $a_1$, $a_2$, and $a_3$ are constants and $\tau$ is related to $t$ by $dt = f_1(\tau)f_2(\tau)^2d\tau$.

\subsubsection{Hamiltonian and generalized momenta}
The Hamiltonian can be written as 
\begin{equation}
    H = \tfrac{N}{\sqrt{\gamma}}H_0 + N^{\alpha}H_{\alpha},
\end{equation}
where
\begin{equation}\label{eq. for bianchi}
    \begin{aligned}
        H_{0} = L_{\alpha\beta\mu\nu}\pi^{\alpha\beta}\pi^{\mu\nu}+ \gamma R, \quad H_{\alpha} = C_{\alpha \rho}^{\mu}\gamma_{\beta \mu}\pi^{\beta \rho}
    \end{aligned}
\end{equation}
with $R = C_{\mu\kappa}^{\alpha}C_{\nu\lambda}^{\beta}\gamma_{\alpha\beta}\gamma^{\mu\nu}\gamma^{\kappa\lambda}$. The generalized momenta are defined similar to the that of Bianchi I.

\subsubsection{Conditional symmetries}
The CSs are generated by $E_{(I)}{}_{\alpha\beta} = \epsilon^{\mu}_{(I)}{}_{\beta}\gamma_{\alpha\mu}$, where $I=1,2,\cdots,6$, and $\epsilon_{(I)}$ are matrices that generate matrices of the form
\begin{equation}\label{gen for [3,3,3]}
    \begin{pmatrix}
        \kappa+\mu &0&0\\
        0&\kappa&\rho\\
        0&\sigma&\mu
    \end{pmatrix}.
\end{equation}
The COMs are given by $Q_{(I)} = E_{(I)}{}_{\alpha\beta}\pi^{\alpha\beta}$. The algebra satisfied by $Q_{(I)}$ is
\begin{equation}\label{COMs of Bianchi II}
    \{Q_{(I)},Q_{(J)}\}= \tilde{C}_{IJ}^MQ_{(M)}, \quad \{Q_{(I)},H_{\alpha}\}=-\tfrac{1}{2}\lambda_{(I)}{}_{\alpha}^{\beta}H_{\beta}, \quad \{Q_{(I)},H_0\} = -2(\kappa+\mu)\gamma R,
\end{equation}
where $\lambda_{(I)}$ are the generators of matrices of the form
\begin{equation}
  \begin{pmatrix}
        \kappa+\mu &x&y\\
        0&\kappa&\rho\\
        0&\sigma&\mu
    \end{pmatrix}.  
\end{equation}
To obtain true COMs, we have to restrict ourselves to the subalgebra where ${\kappa+\mu = 0}$ rendering $\{Q_{(I)},H_0\}=0$. Let us call the generators of this subalgebra $\tilde{E}_{(I)}{}_{\alpha\beta}$ and the COMs $\tilde{Q}_{(I)} = \tilde{E}_{(I)}{}_{\alpha\beta}\pi^{\alpha\beta}$, i.e., the COM algebra is $5-$dimensional and non-Abelian.

\subsubsection{Quantization}

Quantizing the system in the same way as in the last subsection, we get the eigenvalue equations for the CSs
\begin{equation}
    \epsilon_{(I)}^{\alpha}{}_{\beta}\gamma_{\alpha\rho}\tfrac{\partial \Psi}{\partial \gamma_{\beta\rho}}=0,
\end{equation}
where the eigenvalues are taken to be zero in order for $\tilde{Q}_{(I)}$ to generate symmetries \cite{Christodoulakis:2000xe}.
Solving these equations while recognizing that $\epsilon_{(I)}^{\alpha}{}_{\alpha}=0$, we get
\begin{equation}\label{WF for Bianchi II}
    \Psi = \Psi(\gamma,q),
\end{equation}
where $q=C_{\mu\kappa}^{\alpha}C^{\beta}_{\nu\lambda}\gamma_{\alpha\beta}\gamma^{\mu\nu}\gamma^{\kappa\lambda}$. Using \eqref{H_0 operator}, the WDW equation is given by
\begin{equation}
     -\tfrac{1}{2}\left(L_{\alpha\beta\mu\nu}\tfrac{\partial^2\Psi}{\partial \gamma_{\alpha\beta}\partial\gamma_{\mu\nu}}+3\gamma_{\kappa\lambda}\tfrac{\partial \Psi}{\partial \gamma_{\kappa\lambda}}+ \tfrac{3}{2}\Psi\right) + \gamma R \Psi=0.
\end{equation}
This equation can not be solved analytically, i.e., we can only find approximate numerical solutions, but we can impose the CSs to simplify it.  Substituting \eqref{WF for Bianchi II}, we get
\begin{equation}
5q^{2}\tfrac{\partial^{2}\Psi}{\partial q^{2}}
- 3\gamma^{2}\tfrac{\partial^{2}\Psi}{\partial \gamma^{2}}
+ 2q\gamma\tfrac{\partial^{2}\Psi}{\partial \gamma \partial q}
+ 5q\tfrac{\partial \Psi}{\partial q}
- 3\gamma\tfrac{\partial \Psi}{\partial \gamma}
- 2q\gamma\Psi = 0,
\end{equation}
whose solution reads
\begin{equation}\label{WF for [3,3,3]}
    \Psi(\gamma,q) = c_1 I_0\left(\sqrt{2q\gamma}\right)+c_2 K_0\left(\sqrt{2q\gamma}\right),
\end{equation}
where $c_1$ and $c_2$ are constants.

\subsection{Bianchi types VIII and IX --- $[3,3,\{8,9\}]$}
The Killing vectors Bianchi type VIII ($[3,3,8]$) are
\begin{equation}
    \begin{aligned}
        X_1 &= \csch x^3\sin x^4\partial_{x^1}+\cos x^4\partial_{x^3}-\coth x^3\sin x^4\partial_{x^4},\\
        X_2 &= -\cos x^4\csch x^3\partial_{x^1}-\sin x^4\partial_{x^3}-\cos x^4\coth x^3\partial_{x^4}, \quad X_3 = \partial_{x^4},
    \end{aligned}
\end{equation}
while for Bianchi type IX ($[3,3,9]$) they read
\begin{equation}
    \begin{aligned}
        X_1 &= \csc x^3\sin x^4\partial_{x^1}+\cos x^4\partial_{x^3}-\cot x^3\sin x^4\partial_{x^4},\\
        X_2 &= -\cos x^4\csc x^3\partial_{x^1}-\sin x^4\partial_{x^3}-\cos x^4\cot x^3\partial_{x^4}, \quad X_3 = \partial_{x^4}.
    \end{aligned}
\end{equation}
The symmetry-invariant metrics are given by \eqref{eq:Bianchimetric} but with the 1-forms $\sigma^{\alpha}$ satisfying $d\sigma^{\alpha} = \tfrac{1}{2}C_{\beta\gamma}^{\alpha} \sigma^{\beta}\wedge \sigma^{\gamma}$, i.e.,
\begin{equation}
   \sigma^{1} = dx^1-\cosh x^3 dx^4, \quad \sigma^2=\sin x^1 dx^3 + \cos x^1 \sinh x^3 dx^4, \quad \sigma^3 = \cos x^1 dx^3 - \sin x^1 \sinh x^3 dx^4,     
\end{equation}
for Bianchi VIII, and 
\begin{equation}
        \sigma^1= dx^1-\cos x^3 dx^4, \quad \sigma^2=\sin x^1 dx^3 + \cos x^1 \sin x^3 dx^4, \quad \sigma^3 = \cos x^1 dx^3 - \sin x^1 \sin x^3 dx^4,
\end{equation}
for Bianchi IX.
 The non-zero structure constants are, respectively, $C_{23}^1=C_{31}^2 = -C_{12}^3=1$ and $C_{23}^1=C_{31}^2 = C_{12}^3=1$ \cite{Ellis:1968vb}. 
Following the same method as in the previous sections, we get a WDW equation of the same form as \eqref{eq. for bianchi}. For Bianchi type I and II metrics, the WDW equation could only be solved with the help of a large number of COMs or, in other words, many CSs. This does not apply to Bianchi type VIII ($[3,3,8]$) and Bianchi type IX ($[3,3,9]$) where there are no CSs \cite{Ashtekar:1992np}; thus, in the most general case, the equations can not be solved analytically. It is important to stress that this is not a mathematical inconsistency in the minisuperspaces of Bianchi type VIII and IX models, since numerical solutions can be derived.
The quantization of LRS Bianchi type VIII and IX models was done in \cite{Karagiorgos:2018gkn}; the wave functions for the LRS Bianchi VIII/IX are identical to those of s./h. Taub--NUT inside the horizon. Adding a cosmological constant, an aligned electromagnetic field (with the electric and magnetic components being parallel), and a scalar field was discussed in \cite{Berkowitz:2021vho,Paliathanasis:2016rho,Zampeli:2015ojr} and the quantization in a thermal bath with a phenomenological energy density in \cite{Montani:2004hw}. 

\subsection{Extra symmetries}\label{special cases}

As noted in \cite{Frausto:2024egp}, adding Killing vectors (imposing extra symmetries) to $[3,3,-]$ enhances the symmetry to $[4,3,-]$ or even $[6,3,-]$. This can be seen as restricting ourselves to special cases of the Bianchi class A with extra symmetries. For example, axially-symmetric Bianchi type VIII and Bianchi type IX (whose quantization was carried out in \cite{Karagiorgos:2018gkn}) correspond to $[4,3,2]$ and $[4,3,4]$, respectively. In this subsection, we schematically list all such mappings for $[3,3,\{2,3,8,9\}]$ with the Killing vectors to be added, and the coordinate transformations (if applicable) in each case. All these, together with the specialization of the symmetry-invariant metric, can be found in Tab.~X in \cite{Frausto:2024egp}.

\subsubsection{$[4,3,11]$, $[4,3,6]$, $[6,3,2]$ within $[3,3,2]$}
$[3,3,2] \to$ 
\begin{itemize}
    \item $[4,3,11]$: $x^2\partial_{x^1}+x^1\partial_{x^2}$. 
    \item $[4,3,6]$: $-x^4\partial_t+t\partial_{x^4}$, with renaming $t \to x^2$ and $x^2 \to t$.
    \item $[6,3,2]$: $t\partial_{x^2}-x^2\partial_t$, $x^3\partial_{x^2}-x^2\partial_{x^3}$, $x^3\partial_t-t\partial_{x^3}$, with renaming $t \to x^1$ and $x^1 \to t$. 
\end{itemize}

\subsubsection{$[4,3,5]$, $[4,3,10]$ within $[3,3,3]$}
$[3,3,3] \to$
\begin{itemize}
    \item $[4,3,5]$: $x^3\partial_{x^1}-x^1\partial_{x^3}-\tfrac{1}{2}\left((x^1)^2-(x^3)^2\right)\partial_{x^2}$. 
    \item $[4,3,10]$: $x^1\partial_{x^1}-x^3\partial_{x^3}$. 
\end{itemize}

\subsubsection{$[4,3,2]$, $[4,3,9]$ within $[3,3,8]$}
$[3,3,8] \to $ 
\begin{itemize}
    \item $[4,3,2]$: $\partial_{x^1}$; the quantized system matches that of $[4,3,2]$ only inside the horizon.
    \item  $[4,3,9]$: $\partial_{x^2}$, with transformations: $\tan x^1 \to \coth t\sinh x^2$,  $t \to x^3$, $\sech x^2 \to \sech t\sech x^2$,
            $\coth x^3 \to \tfrac{1-\tan x^1\csch t\tanh x^2}{\tan x^1+\csch t\tanh x^2}$.
\end{itemize}

\subsubsection{$[4,3,4]$ within $[3,3,9]$}

$[3,3,9] \to $ 
\begin{itemize}
     \item $[4,3,4]$: $\partial_{x^1}$; the quantized system matches that of $[4,3,4]$ only inside the horizon.
\end{itemize}

\section{Conclusions}\label{Summary and conclusions}

In this paper, we carried out the minisuperspace quantization procedure for all PSC-compatible models classified via the Hicks classification. We reviewed the minisuperspace quantization of the s. Schwarzschild, FLRW, and Bianchi type I, II, VIII, IX spacetimes, and presented for the first time the quantizations of h./p. Schwarzschild, s./h./p. Taub–NUT, BI/BII/BIII metrics, the swirling universe, near-horizon extremal Kerr, and $\mathrm{AdS_3}$/$\mathrm{M_3}$/$\mathrm{dS_3}$ homogeneous spacetimes. Note that the regions inside the horizons of the s./h. Taub–NUT spacetimes are special cases of the Bianchi IX and VIII models, respectively, which have been previously derived in the literature. However, here, we derived the wave function that is valid inside and outside the horizon. In all cases, we systematically derived the reduced Lagrangians and the generalized momenta, thus, defining the corresponding minisuperspaces. We then computed the algebras of CSs and the COMs they induce. Finally, we applied the minisuperspace quantization procedure, promoting the coordinates and the generalized momenta to operators, writing and solving their WDW equations. In vacuum, a summary of the results obtained in the paper is displayed in Tab.~\ref{table 1}.

To obtain nontrivial solutions within $[6,3,-]$, we introduced a cosmological constant and a minimally coupled massless scalar field, since otherwise the only symmetry-invariant solutions are Minkowski spacetime (for flat FLRW) or Milne spacetime (for open FLRW). A summary of their CSs and quantization is displayed in Tab.~\ref{table2}. Similarly, for $[3,3,-]$, the only symmetry-invariant metrics admitting vacuum solutions are for Bianchi type I ($[3,3,2]$) and II ($[3,3,3]$) symmetries. On the other hand, Bianchi types VIII ($[3,3,8]$) and IX ($[3,3,9]$) admit only closed form exact vacuum solutions with enhanced symmetries, and thus are not strictly within $[3,3,-]$.

A potential concern regarding the results obtained in this paper is that we relied on PSC to ensure that the equations derived from the symmetry-reduced theory coincide with the reduced equations of the full theory. While PSC guarantees this equivalence for any gravitational theory, one might still wonder whether PSC-incompatible models could nevertheless exhibit such an equivalence in a specific theory, e.g. GR. However, we have verified that this is not the case for all PSC-incompatible metrics in GR by directly comparing the reduced Einstein equations with the Euler–Lagrange equations derived from the corresponding reduced Lagrangians, and finding that they differ.

While PSC guarantees the equivalence of the field equations between the full and reduced models at the classical level, this is less clear quantum mechanically, since there is no quantum analogue of PSC. The reason is that we only have access to the symmetry-reduced theories and their associated wave functions. The corresponding quantum description would be given by the full quantum gravity equations, which are infinite-dimensional: the WDW equation is a functional equation that is generally ill-defined (at best, only formally defined as an infinite system of coupled differential equations). Another possible route to derive the full quantum dynamics and test this equivalence would be loop quantum gravity; however, this programme is still incomplete, and the problem remains open.

Our results naturally motivate future work aimed at extracting physical interpretations from the CSs and wave functions of minisuperspace models, in particular probabilistic statements such as the likelihood that a quantum solution admits curvature singularities, is asymptotically flat, possesses a horizon, or has a regular symmetry axis. The main difficulty is that the measure on minisuperspace cannot be fixed by the quantization procedure alone, so the probability distribution remains undetermined, since the scheme does not select a unique supermetric and superpotential but rather a class related by the rescalings discussed in Sec.~\ref{Quantization of $[4,3,-]$}. One must therefore either fix a specific supermetric and superpotential or impose additional conditions to determine the measure. 

A common method to fix the probability distribution is to demand that all constants of motion operators be Hermitian, which only works if there are more COMs than dimensions of the minisuperspace. However, in this paper, we only impose Hermiticity on a COM when using its eigenvalue equation as a constraint in addition to the WDW equation. A geometrically motivated way to select a supermetric and a superpotential is to rescale them such that the superpotential is constant. In this case, the CS generators are Killing vectors of the supermetric (instead of conformal Killing vectors), and the CSs can be identified with Noether symmetries. This fixes the measure, up to a constant, to the square root of the absolute value of the determinant of the supermetric for which the superpotential is constant. This works without conditions on the COM operators, so it is valid for minisuperspaces with no CSs, such as the s./h. Taub--NUT cases. Other conditions may also be imposed; for example, requiring the probability distribution to be Gaussian, or normalizable in general, may be useful when comparing with semiclassical counterparts. However, an agreed-upon rigorous way to fix the probability distribution (up to a constant) is still lacking.

Another application to be considered is the calculation of physical quantities such as mass or angular deficits for the quantum spacetimes in question. This has been done in semiclassical settings, where the quantum spacetime is approximated by a Gaussian function peaked at the classical value. This facilitates the derivation of physical quantities similar to those of the classical spacetime \cite{Davidson:2014tda}; however, constructing a general method to extract physical quantities from full quantum theories remains an open problem.

\begin{table}[h]
\centering
\begin{tabular}{|c|c|c|c|}
\hline
Hicks \# &
\makecell{CSs} &
Algebra of COMs &
Wave function \\
\hline
\makecell{
$[4,3,1]$ (h. Schwarzschild)\\
$[4,3,3]$ (s. Schwarzschild)\\
$[4,3,8]$ (BI/II-metric)}&\makecell{
$\xi_1=(-q_1,q_4)$, $\xi_2=\left(\tfrac{1}{q_4},0\right)$,\\
$\xi_3=\left(-\tfrac{q_1}{2\sqrt{q_4}},\sqrt{q_4}\right)$}& \makecell{$\{Q_1,Q_3\}=Q_3$, $\{Q_2,Q_1\}=Q_2$,\\ $\{Q_2,Q_3\}=0$\\(Bianchi VI$_{-1}$)}&
\makecell{\eqref{WF for [4,3,3]}\\ with $k=\pm1$}
\\ \hline
\makecell{$[4,3,2]$ (h. Taub--NUT)\\
$[4,3,4]$ (s. Taub--NUT)\\$[4,3,9]$ (NHEK)}& No CSs & No COMs& \makecell{\eqref{WF for [4,3,(4,2)]}\\ with $k=\pm 1$}
\\ \hline
\makecell{$[4,3,5]$ (p. Taub--NUT)\\
$[4,3,10]$ (swirling universe)}&\makecell{
$\xi_1=(0,q_4)$, $\xi_2=\left(\tfrac{1}{q_1q_4},0\right)$,\\
$\xi_3=\left(q_1q_4,-2q_4^2\right)$}&\makecell{
$\{Q_1,Q_2\}=Q_2$,\\
$\{Q_1,Q_3\}=-Q_3$, $\{Q_2,Q_3\}=0$\\(Bianchi VI$_{-1}$)}&
\eqref{WF for [4,3,5]}
\\ \hline
\makecell{$[4,3,6]$ (p. Schwarzschild)\\
$[4,3,11]$ (BIII-metric)}&\makecell{
$\xi=(f,g)$ with\\$f,g$ given by\\
\eqref{f function} and \eqref{g function}}& \makecell{Infinite-dimensional}&
\eqref{WF for [4,3,6]}
\\ \hline
\makecell{$[6,3,1]$ (closed FLRW)\\
$[6,3,2]$ (flat FLRW)\\
$[6,3,3]$ (open FLRW)\\
$[6,3,4]$ ($\mathrm{AdS_3}$-homogeneous)\\
$[6,3,5]$ ($\mathrm{M_3}$-homogeneous)\\
$[6,3,6]$ ($\mathrm{dS_3}$-homogeneous)}&$\xi = 1$& \makecell{$1d$ Abelian (Trivial)}&\makecell{\eqref{WF of [6,3,{1,3}]} for ${k=\pm1}$\\ \eqref{WF for [6,3,2]} for ${k=0}$}
\\ \hline
\makecell{$[3,3,2]$ (Bianchi type I)}&\makecell{
$E_{(I)\beta\rho}= \lambda_{(I)\beta}^{\alpha}\gamma_{\alpha\rho}$,\\
$\lambda_{(I)\beta}^{\alpha}$ given by\\
\eqref{generators for [3,3,1]}}&\makecell{$9d$ non-Abelian with\\
$\{Q_{(I)},Q_{(J)}\}= -\tfrac12 C_{IJ}^M Q_{(M)}$,\\
$C_{IJ}^M$ given by\\\eqref{structure constants for [3,3,1]}}&\eqref{WF for [3,3,1]}
\\ \hline
\makecell{$[3,3,3]$ (Bianchi type II)}&
\makecell{$E_{(I)\alpha\beta}= \epsilon_{(I)\beta}^{\mu}\gamma_{\alpha\mu}$,\\
$\epsilon_{(I)\beta}^{\mu}$ generate\\\eqref{gen for [3,3,3]}}&\makecell{$5d$ non-Abelian with \eqref{COMs of Bianchi II}\\ and $\kappa + \mu=0$.}
&\eqref{WF for [3,3,3]}
\\ \hline
\makecell{$[3,3,8]$ (Bianchi type VIII)\\
$[3,3,9]$ (Bianchi type IX)}& \makecell{No CSs} & \makecell{No COMs}&
\makecell{
WDW eq. unsolvable;\\ special cases in Sec.~\ref{special cases}
}
\\ \hline
\end{tabular}
\caption{
Summary of PSC-compatible metrics, their CSs,
and the wave functions of the corresponding quantum systems in vacuum. Names in parentheses refer to representative spacetimes with such spacetime symmetries.}
\label{table 1}
\end{table}

\begin{table}[h]
    \centering
    \begin{tabular}{|c|c|c|c|}
    \hline
    Hicks \#& CSs & Algebra of COMs& Wave function \\
    \hline
         \makecell{$[6,3,1]$,$[6,3,3]$,\\$[6,3,4]$,$[6,3,6]$}& $\xi=(0,1)$ & \makecell{$1d$ Abelian with $Q=p_{\phi}$} & \eqref{WF for [6,3,1] with matter} with $k=\pm 1$ \\
         \hline
         \makecell{$[6,3,2]$,$[6,3,5]$}& \makecell{ $\xi_1=(0,1)$, $\xi_2=\left(\tfrac{e^{-\sqrt{3/2}\phi}}{q_2^2},\sqrt{6}\tfrac{e^{-\sqrt{3/2}\phi}}{q_2^3}\right)$\\ $\xi_3=\left(\tfrac{e^{\sqrt{3/2}\phi}}{q_2^2},-\sqrt{6}\tfrac{e^{\sqrt{3/2}\phi}}{q_2^3}\right)$} & \makecell{$\{Q_1,Q_2\}=Q_2$, $\{Q_3,Q_1\}=Q_3$,\\ $\{Q_3,Q_2\}=0$\\ (Bianchi VI$_{-1}$)}  & \eqref{WF for [6,3,2] with matter} \\
         \hline
    \end{tabular}
    \caption{Summary of CSs and wave functions for FLRW models with a cosmological constant and a massless scalar field. 
    }
    \label{table2}
\end{table}

\begin{acknowledgments}
I.K. and P.T. acknowledge financial support from the Primus grant PRIMUS/23/SCI/005 of Charles University. I.K. further thanks the Charles University Research Center grant UNCE24/SCI/016 for support. P.T. acknowledges financial support by GAUK 425425 from Charles University.
\end{acknowledgments}

\end{document}